%% file: main.tex
\documentclass[10pt,journal,compsoc]{IEEEtran}

\input{macros}

\newmdenv[
  topline=false,
  bottomline=false,
  rightline=false,
  leftline=true,
  linewidth=0.8pt,
  linecolor=black,
  innertopmargin=2pt,
  innerbottommargin=2pt,
  innerleftmargin=8pt,
  innerrightmargin=0pt,
  skipabove=6pt,
  skipbelow=6pt
]{rqanswer}

\title{When LLM-Generated Refactorings Ripple: An Empirical Study of Repository Adoption Readiness}

\author{Muhammad Waseem, Aakash Ahmad, and Pekka Abrahamsson%
\IEEEcompsocitemizethanks{
\IEEEcompsocthanksitem Muhammad Waseem and Pekka Abrahamsson are with Tampere University, Finland.
E-mail: \{muhammad.waseem, pekka.abrahamsson\}@tuni.fi.
\IEEEcompsocthanksitem Aakash Ahmad is with the School of Computing and Engineering, University of Derby, UK.
E-mail: a.abbasi@derby.ac.uk.
}}

\begin{document}

\IEEEtitleabstractindextext{%

\begin{abstract}
Large Language Models (LLMs) are increasingly used to refactor existing source files to improve their structure and maintainability. Existing evaluations mainly measure file-local outcomes, such as smell reduction, compilation, or test success. However, these outcomes do not reveal whether a candidate changes contracts, dependencies, or caller assumptions that affect its adoption within the surrounding repository. To address this gap, we conduct 1,500 generation attempts across 150 production files from 15 open-source Java repositories, five LLM configurations, and two dependency-aware generation conditions: single-prompt refactoring and LangGraph-based multi-agent refactoring. Of these attempts, 1,455 yield complete candidates eligible for analysis. We evaluate these candidates using local changes, repository-facing structural indicators, detected caller impact, the behavioural-equivalence indicator, and available test evidence, and categorize them as \textsc{Low-risk}, \textsc{Review-needed}, \textsc{High Contract/Dependency Risk}, or \textsc{Inconclusive}. The results show that 1,365 candidates (93.8\%) are classified as having High contract/dependency risk, whereas 106 (7.3\%) have at least one detected broken downstream caller and 290 (19.9\%) have a false behavioural-equivalence indicator; only eight satisfy the Low-risk criteria. Within the reason-covered subset, the High-risk rate decreases under more conservative dependency-delta thresholds, while 60 of the 65 candidates with conclusive targeted-test outcomes pass, including 50 classified as High risk. Paired analyses of 705 complete pairs identify no statistically significant generation-condition difference in the principal risk indicators, although model configurations differ in specific detected risk channels. These findings suggest that practitioners should treat LLM-generated refactorings as candidates requiring staged repository-level verification, while future research should validate screening and extraction precision and strengthen compilation, behavioural, and coordinated multi-file verification.
\end{abstract}

}

\maketitle

\IEEEdisplaynontitleabstractindextext

\IEEEpeerreviewmaketitle

\input{sections/01-introduction}
\input{sections/02-background-related-work}
\input{sections/03-methodology}
\input{sections/04-results-and-analysis}

\input{sections/05-discussion}

\input{sections/06-threats-to-validity}

\input{sections/07-conclusion}

\section*{Acknowledgments}

During the preparation of this work, the authors used OpenAI's
ChatGPT\footnote{\url{https://chatgpt.com/}} to assist with
language editing and polishing throughout the manuscript.
Claude's Excel extension\footnote{\url{https://claude.ai/}}
was used to help organize, format, and polish the accompanying
Excel data sheets. The authors also used
Cursor\footnote{\url{https://cursor.com/}} to assist with
implementing the experimental software described in Section~3.
The authors reviewed and revised the AI-assisted text,
spreadsheet changes, and code, and take full responsibility
for the final manuscript, dataset, software, analyses, and
conclusions.

\section*{Data Availability}

The data and source code used in this study are available at \cite{rippleref_data} and \cite{rippleref_code}, respectively.

\bibliographystyle{IEEEtran}
\bibliography{References}

\appendices

\end{document}

%% file: macros.tex
\usepackage[utf8]{inputenc}
\usepackage[T1]{fontenc}
\usepackage{amsmath,amssymb}
\usepackage{amsthm}
\usepackage{graphicx}
\usepackage{booktabs}
\usepackage{array}
\usepackage{multirow}
\usepackage{url}
\usepackage{xspace}
\usepackage{enumitem}
\usepackage{amsmath}
\usepackage{amssymb}
\usepackage{hyperref}
\usepackage{booktabs}
\usepackage{multirow}
\usepackage{amsmath}
\usepackage{amssymb}
\usepackage{array}
\usepackage{booktabs}
\usepackage{multirow}
\usepackage{graphicx}
\usepackage[most]{tcolorbox}
\usepackage{tabularx}
\usepackage{array}
\usepackage{placeins}
\usepackage[table]{xcolor}
\usepackage{fontawesome5}
\usepackage{mdframed}

\hypersetup{
    colorlinks=true,
    linkcolor=black,
    citecolor=black,
    urlcolor=black
}

\newcolumntype{L}[1]{>{\raggedright\arraybackslash}p{#1}}
\newcolumntype{C}[1]{>{\centering\arraybackslash}p{#1}}
\newcolumntype{R}[1]{>{\raggedleft\arraybackslash}p{#1}}

\setlist[itemize]{leftmargin=*, noitemsep, topsep=2pt}
\setlist[enumerate]{leftmargin=*, noitemsep, topsep=2pt}

\graphicspath{{figures/}}

%% file: sections/01-introduction.tex
\section{Introduction}
\label{sec:introduction}

Refactoring improves the internal structure of software while preserving its externally observable behaviour~\cite{fowler2018refactoring,mens2004survey}. It can improve maintainability and facilitate future changes, but an incorrect transformation may introduce regressions, integration problems, or unintended behavioural changes~\cite{kim2012field,murphyhill2012how}. Large Language Models (LLMs) are increasingly used to refactor existing source files by interpreting code-quality problems and generating context-sensitive revisions~\cite{hou2024large}. Recent empirical studies show that LLMs can produce useful refactoring suggestions, reduce detected code smells, and improve file-local quality indicators in some settings~\cite{depalma2024exploring,cordeiro2026capability}. Unlike a mechanically specified refactoring operation, however, an LLM-generated output is a candidate whose suitability for repository integration must be established before adoption.

The central problem is that evidence of improvement within the edited file does not show whether the candidate preserves the properties on which the surrounding repository depends. Java files expose methods and constructors to callers, depend on imported types, implement interfaces, participate in inheritance structures, and use annotations that may control framework behaviour. A change to any of these elements can affect code beyond the edited file, even when the candidate reduces local smells or appears structurally cleaner. Repository-level code-generation research similarly emphasizes that realistic changes require context from multiple files and must account for dependencies and constraints distributed throughout a repository~\cite{bairi2024codeplan}.

\begin{figure*}[t]
\centering
\includegraphics[width=\textwidth]{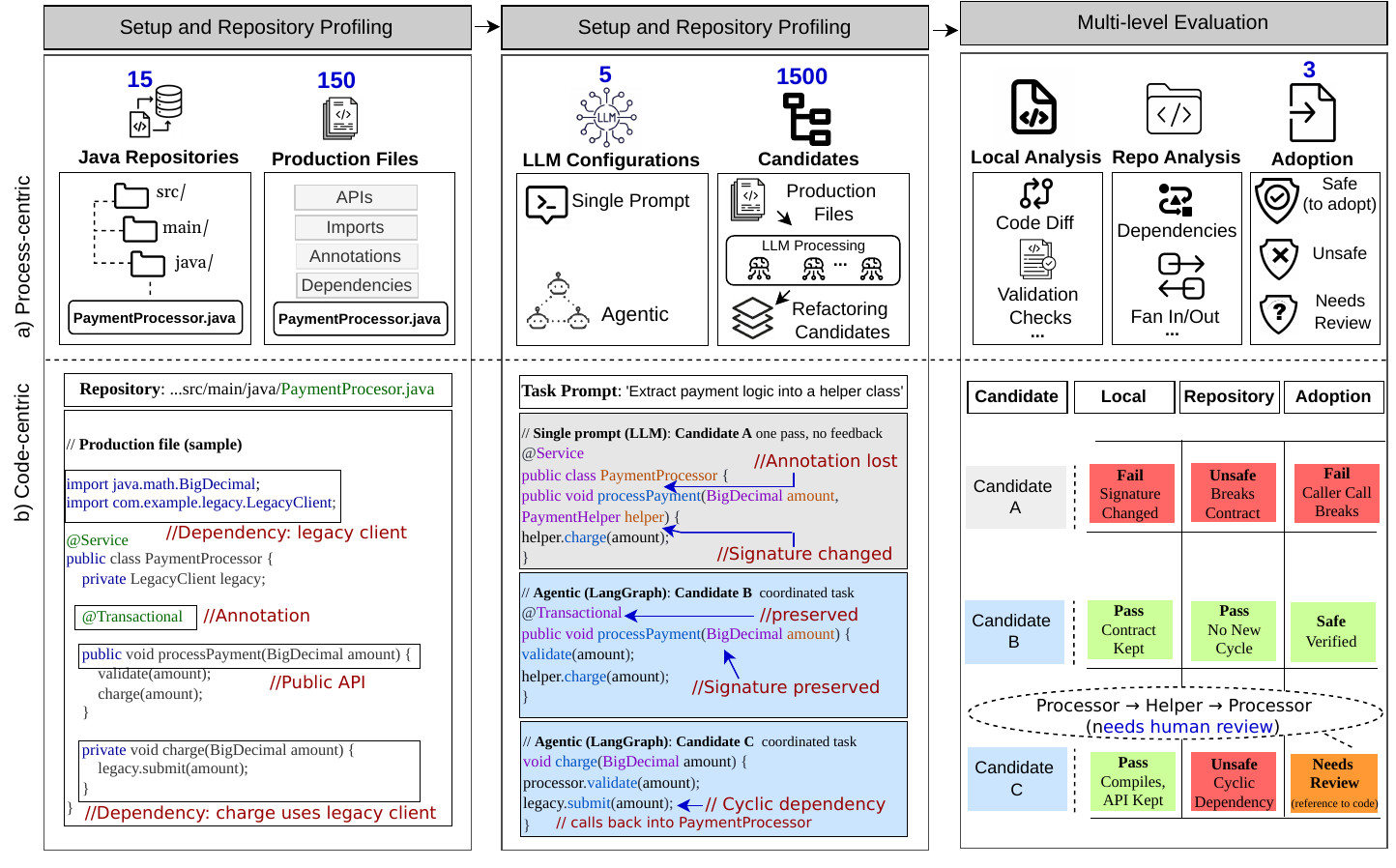}
\caption{Local improvement and repository-level verification of an LLM-generated refactoring candidate. A candidate may improve the edited file while changing a public contract, caller-used API, dependency relation, framework-sensitive construct, or downstream caller assumption.}
\label{fig:intro-local-vs-repository-risk}
\end{figure*}

Consider a candidate that decomposes a long method and removes a locally detected smell but also changes the signature of a public method or constructor. The edited file may appear improved, while existing callers can no longer invoke it correctly. Similarly, changing an annotation, implemented interface, or dependency may alter how the class is instantiated, discovered, or used elsewhere. Figure~\ref{fig:intro-local-vs-repository-risk} illustrates this distinction between file-local improvement and repository-level verification. A structural warning does not by itself demonstrate semantic failure, but it identifies a repository-facing change that may require inspection, coordinated modification, compilation, or testing before adoption.

Existing studies provide important evidence about LLM refactoring capability, local quality improvement, compilation, testing, and behaviour preservation in controlled settings~\cite{depalma2024exploring,cordeiro2026capability}. Repository-aware and multi-agent workflows further use retrieval, specialized roles, and tool feedback to improve candidate generation and verification~\cite{chondamrongkul2026repoai,xu2025mantra,oueslati2026refagent}. Nevertheless, limited evidence is available about the repository-facing properties that remain in generated candidates. In particular, it remains unclear how frequently candidates alter contracts and dependencies, how often these changes coincide with detected caller or behavioural impact, and whether single-prompt and multi-agent generation produce different risk profiles when evaluated on the same files and models. This gap matters because adoption decisions require developers to assess not only whether a file improves, but also whether the resulting change can be reviewed, integrated, compiled, tested, and trusted within the surrounding system.

To examine this gap, we conduct an empirical study of LLM-generated Java refactoring candidates. The study covers 15 open-source repositories and 150 production files selected to combine local refactoring need with observable repository exposure. For each selected file, we attempt candidate generation with five LLM configurations under two dependency-aware conditions: single-prompt refactoring and LangGraph-based multi-agent refactoring. This design produces 1,500 generation attempts, of which 1,455 yield complete candidates eligible for analysis, comprising 706 single-prompt and 749 multi-agent candidates. The two conditions receive the same selected files and repository dependency context, enabling matched comparisons across 705 complete file--model pairs.

We evaluate each eligible candidate using complementary forms of evidence. File-local measures characterize the generated edit, while repository-facing indicators capture changes to public and protected contracts, caller-used APIs, dependencies, annotations, inheritance relations, implemented interfaces, and detected downstream callers. The behavioural-equivalence indicator and available targeted-test outcomes provide additional evidence about behaviour preservation. Based on the evidence produced by the verification pipeline, candidates are categorized as \emph{Low-risk}, \emph{Review-needed}, \emph{High contract/dependency risk}, or \emph{Inconclusive}. These categories support candidate screening and review prioritization; they are not treated as independently validated judgments of semantic correctness. We further compare the generation conditions and model configurations and examine exploratory associations between source-file exposure, generated changes, and the resulting screening and observed-impact outcomes. The research questions are presented in Section~\ref{subsec:objective-rqs}.

This paper makes the following contributions.

\begin{itemize}
    \item We formulate and operationalize a repository-level evaluation perspective for LLM-generated refactoring that distinguishes file-local improvement, structural screening, detected caller impact, and behavioural verification evidence.

    \item We conduct a controlled study of 1,500 generation attempts across 15 Java repositories, 150 production files, five LLM configurations, and two dependency-aware generation conditions. To the best of our knowledge, this is the first paired empirical study examining whether multi-agent orchestration changes repository-level ripple-risk and verification evidence under matched file and model conditions.

    \item We provide the generation and verification implementation, raw study records, final analysis dataset  \cite{rippleref_data}, and analysis scripts  \cite{rippleref_code} to support reproduction of the quantitative analyses and artifact-level inspection where the required candidate artifacts are available.
\end{itemize}

The remainder of this paper is organized as follows. Section~\ref{sec:background-related-work} reviews related work. Section~\ref{sec:methodology} describes the study design and analysis procedures. Section~\ref{sec:results} presents the empirical results, and Section~\ref{sec:discussion} discusses their implications. Section~\ref{sec:threats} examines threats to validity, and Section~\ref{sec:conclusion} concludes the paper.

%% file: sections/02-background-related-work.tex
\section{Background and Related Work}
\label{sec:background-related-work}

This section reviews prior work most directly related to our study, including empirical evidence on LLM-based refactoring, repository-aware and multi-agent refactoring workflows, and verification-oriented refactoring benchmarks. Rather than providing a general introduction to refactoring, we focus on the evidence needed to position our investigation of repository-level risk screening and adoption readiness.

\subsection{Effectiveness and Limitations of LLM-Based Refactoring}
\label{subsec:rw-llm-refactoring}

Empirical studies show that LLMs can support code refactoring, but their effectiveness depends on task scope, prompt design, model choice, and the evidence used to evaluate the generated changes. Shirafuji et al.~\cite{shirafuji2023refactoring} studied few-shot LLM-based refactoring on short Python programs and reported reductions in cyclomatic complexity and lines of code among semantically valid outputs. Their findings demonstrated the potential of example-guided generation while also revealing unintended changes, such as the removal or translation of comments.

Liu et al.~\cite{liu2025exploring} evaluated general-purpose LLMs on real Java refactorings. They found that prompts specifying the expected refactoring category substantially improved the identification of refactoring opportunities. However, some recommended solutions changed program functionality or introduced syntax errors. Their results show that constrained prompting can improve task performance but does not eliminate the need for subsequent verification.

Cordeiro et al.~\cite{cordeiro2026capability} evaluated multiple LLM configurations on refactoring tasks drawn from 30 open-source Java projects. They examined code-smell reduction, automated quality indicators, and unit-test outcomes under different prompting strategies. The results showed that LLMs can reduce detected smells and successfully perform some multi-file refactorings, although effectiveness varies across models, refactoring types, and prompting strategies.

Other studies further demonstrate that behaviour preservation and collateral effects require explicit evaluation. Zhang et al.~\cite{zhang2026consistency} examined behavioural consistency across Java and Python refactoring outputs and found that behaviour preservation cannot be assumed merely because a generated transformation is presented as a refactoring. Freitas et al.~\cite{freitas2026evaluating} studied LLM-assisted Java refactoring from a developer perspective and found that smell-targeted changes can introduce collateral quality issues, including new smell instances. Cordeiro et al.~\cite{cordeiro2025limitations} similarly identified hallucination, limited project context, semantic uncertainty, and the continued need for human verification as barriers to dependable LLM-based refactoring.

Collectively, these studies establish that LLMs can generate useful refactoring candidates, particularly when prompts provide explicit objectives and constraints. However, they also show that local improvement, smell reduction, or task completion does not independently establish behaviour preservation or readiness for repository integration.

\subsection{Repository-Aware and Multi-Agent Refactoring}
\label{subsec:rw-agentic-refactoring}

Recent work has addressed the limited context and verification capabilities of direct LLM prompting by incorporating repository information, expert tools, retrieval mechanisms, and multi-agent orchestration.

Wu et al.~\cite{wu2024ismell} proposed iSMELL, which combines multiple expert code-smell detection tools through a mixture-of-experts architecture and supplies their outputs to LLMs for refactoring. The study demonstrates that external analysis tools can provide more targeted evidence than relying on an LLM alone for both smell detection and refactoring.

RepoAI~\cite{chondamrongkul2026repoai} combines retrieval-augmented generation with multi-agent orchestration for repository-level refactoring. Its workflow coordinates specialized agents and repository-context retrieval while using cross-family validation to assess generated changes. The study reports that multi-agent coordination and repository retrieval improve refactoring outcomes compared with direct generation and self-refinement baselines.

MANTRA~\cite{xu2025mantra} focuses on method-level Java refactoring. It combines context-aware retrieval, Developer and Reviewer agents, and iterative feedback from compilation and test execution. Its evaluation shows that coordinated agents and tool feedback can substantially improve the production of compilable and test-passing refactorings compared with direct LLM generation.

RefAgent~\cite{oueslati2026refagent} similarly coordinates specialized agents for planning, generation, compilation, testing, and iterative correction. Its evaluation compares the multi-agent workflow with single-agent generation, a search-based refactoring tool, and developer-written refactorings. The results indicate that agent specialization and iterative verification can improve compilation, test, and code-quality outcomes.

These approaches provide important evidence that retrieval, repository context, expert tools, compilation, testing, and specialized agents can improve automated refactoring. However, their primary outcomes are generally compilation success, test success, smell reduction, quality metrics, similarity to developer refactorings, or task-completion rates. These measures assess important aspects of refactoring effectiveness, but they do not necessarily isolate repository-facing changes to public contracts, caller-used APIs, dependency relations, framework-sensitive structures, and downstream callers.

Consequently, repository-aware generation and repository-level risk assessment represent related but distinct concerns. Repository context can help a model generate a more informed candidate, while repository-level assessment examines how the resulting candidate changes structures on which the surrounding repository may depend.

\subsection{Verification-Oriented Refactoring Benchmarks}
\label{subsec:rw-verification-benchmarks}

Recent benchmarks and evaluation frameworks have made LLM-based refactoring evaluation more realistic and multidimensional. SmellBench~\cite{lin2026smellbench} evaluates code agents on fine-grained refactoring tasks constructed by injecting controlled code smells into code drawn from real repositories. Its evaluation separates functional correctness, smell localization, and refactoring quality, showing that successful execution does not necessarily imply successful smell removal.

Dinu et al.~\cite{dinu2026smellbench} introduced a benchmark for architectural-smell repair that distinguishes repair effectiveness, false-positive identification, and net codebase impact. Their study shows that tool-reported architectural smells may include false positives and that aggressive repair can introduce new smells. These findings demonstrate the importance of evaluating both the intended improvement and the collateral effects of a generated transformation.

SWE-Refactor~\cite{xu2026swerefactor} provides a repository-level benchmark containing developer-written, behaviour-preserving Java refactorings. It covers both atomic and compound transformations and validates benchmark instances using compilation, test execution, and automated refactoring-detection tools. Its results show that compound and context-dependent refactorings remain challenging even when repository information and automated verification are available.

Prior empirical work on the relationship between code smells and refactoring operations further shows that smell removal and refactoring activity do not have a simple one-to-one relationship. Shetty and Sharma~\cite{shetty2025mapping} analyzed code smells and refactorings across Java repositories and found that many applied refactorings do not remove detected smells, while many removed smells cannot be directly associated with a corresponding refactoring operation.

These benchmarks and empirical studies demonstrate the importance of combining multiple forms of evidence. Nevertheless, benchmark success is not identical to candidate-level adoption readiness. A candidate may complete a target transformation, reduce a local smell, compile, or pass available tests while still modifying a public signature, removing a caller-used API, changing dependency structure, or altering a framework-sensitive construct. Repository-level screening therefore complements compilation, testing, and local quality evaluation by making such changes visible before adoption.

\subsection{Positioning of This Study}
\label{subsec:study-positioning}

Prior studies show that LLMs can generate useful refactoring candidates and that their performance can improve through constrained prompting, repository context, retrieval, expert tools, compilation and test feedback, and multi-agent orchestration. Existing work also establishes that generated refactorings can contain syntax errors, behavioural inconsistencies, collateral quality problems, and context-related failures.

Our study does not introduce the first multi-agent refactoring workflow, nor does it treat multi-agent orchestration as sufficient evidence of repository-level safety. Instead, it provides a matched empirical comparison of dependency-aware single-prompt and LangGraph-based multi-agent generation under the same selected files, model configurations, repository context, and screening implementation.

The study contains 1,500 generation attempts across 150 production files from 15 open-source Java repositories and five LLM configurations. Of these attempts, 1,455 produce eligible candidates for candidate-level analysis. The comparison between generation conditions uses 705 complete file--model pairs for which both single-prompt and multi-agent generation produce eligible candidates.

For each eligible candidate, we distinguish local change measures, contract and dependency risk, detected caller impact, behavioural-risk indicators, and available test evidence. This design allows us to examine whether multi-agent orchestration changes particular repository-level risk channels rather than evaluating it only through smell reduction, compilation success, or test outcomes.

The contribution of this study is therefore not another demonstration that LLMs or multi-agent systems can perform refactoring. It is an empirical examination of the repository-facing evidence that remains after candidate generation and of how this evidence varies across matched generation conditions and model configurations. In this way, the study positions repository-level screening as a complement to local quality measurement, compilation, and testing when assessing whether an LLM-generated refactoring candidate is ready for further review or adoption.

%% file: sections/03-methodology.tex
\section{Methodology}
\label{sec:methodology}

\begin{figure*}[t]
    \centering
    \includegraphics[width=1\linewidth]{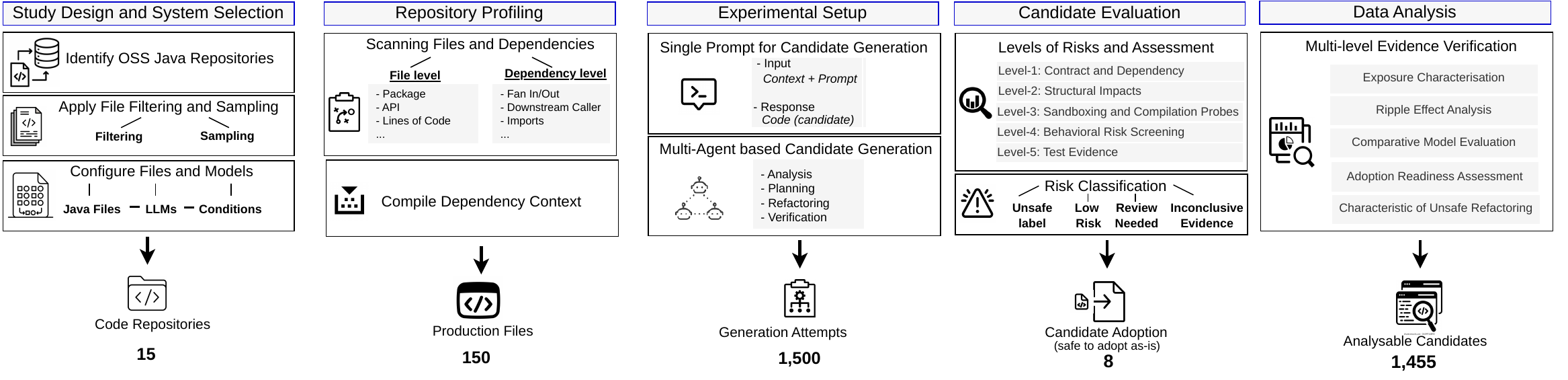}
    \caption{Overview of the study methodology. The study selects production Java files with local refactoring need and repository-level exposure, constructs dependency context, generates candidates under dependency-aware single-prompt and LangGraph-based multi-agent conditions, and evaluates eligible candidates using local change measures, repository-level risk indicators, staged verification evidence, and statistical analyses. Generation completeness is assessed before candidate-level analysis.}
    \label{fig:study-overview}
\end{figure*}

This section describes the empirical design used to study repository-level risk screening and verification evidence for LLM-generated Java refactoring candidates. The study does not treat local code improvement as sufficient evidence of readiness for repository integration. Instead, it examines repository exposure, structural changes, detected caller impact, the behavioural-equivalence indicator, and the compilation and test evidence available for each candidate. Figure~\ref{fig:study-overview} summarizes the study design.

\subsection{Research Objective and Questions}
\label{subsec:objective-rqs}

The objective of this study is to investigate repository-level risk screening and verification evidence for LLM-generated Java refactoring candidates. We examine the repository exposure of the selected files, the ripple-risk indicators detected after generation, variation across generation conditions and model configurations, staged adoption-readiness outcomes, and characteristics associated with contract, dependency, and observed-impact outcomes. We treat these indicators as evidence for candidate screening and review prioritization rather than as proof of semantic failure or production-level safety.

Guided by this objective, we formulate the following five Research Questions (RQs).

\begin{itemize}
    \item \textit{RQ1. What repository-level exposure characterizes the Java files selected for LLM-based refactoring?}

    \item \textit{RQ2. What repository-level ripple-risk indicators are detected in dependency-aware LLM-generated refactoring candidates?}

    \item \textit{RQ3. How do repository-level ripple-risk indicators differ across generation conditions and LLM configurations?}

    \item \textit{RQ4. To what extent do generated candidates satisfy the pipeline's staged adoption-readiness criteria, and what verification evidence supports those screening outcomes?}

    \item \textit{RQ5. Which source-file characteristics and generated-change indicators are associated with the pipeline's contract, dependency, and observed-impact outcomes?}
\end{itemize}

These questions examine repository-level refactoring risk from complementary perspectives. RQ1 characterizes the selected files before generation because repository exposure determines the opportunities for a local refactoring to affect other parts of the system. RQ2 examines the repository-level ripple-risk indicators detected in eligible candidates, including contract and API changes, dependency changes, caller-impact evidence, framework-related changes, and the behavioural-equivalence indicator. RQ3 compares these indicators across the two dependency-aware generation conditions and five LLM configurations using complete eligible pairs and complete model blocks.

RQ4 examines how candidates progress through the pipeline's staged adoption-readiness criteria and reports the structural, caller-impact, behavioural, compilation, and test evidence supporting the resulting screening outcomes. These outcomes are conservative screening categories rather than confirmed correctness or failure judgments. Finally, RQ5 examines exploratory associations between source-file characteristics, generated-change indicators, and three distinct outcomes: pipeline-defined High contract/dependency risk, detected broken-caller evidence, and a false behavioural-equivalence indicator. This separation avoids treating a rule-based classification as an independently validated failure outcome.

\subsection{Study Design}
\label{subsec:study-design}

The experimental design contains 15 open-source Java repositories, 150 selected production Java files, five LLM configurations, and two dependency-aware generation conditions. For each selected file, each model configuration attempts to generate one candidate under dependency-aware single-prompt generation and one under LangGraph-based multi-agent generation. The design therefore contains 1,500 generation attempts:

\begin{equation}
150\ \text{files} \times 5\ \text{models} \times 2\ \text{conditions}
=1{,}500.
\end{equation}

The study distinguishes a recorded generation attempt from an analytically eligible candidate. A generation attempt is eligible for candidate-level analysis only when it yields a complete Java candidate suitable for the configured before--after profiling and verification pipeline. Truncated generations, missing candidate bodies, outputs without the required primary type, and other incomplete or structurally invalid generations are retained in generation-completeness accounting but excluded from candidate-level risk analysis. This distinction prevents incomplete outputs from receiving risk classifications derived from partial code.

Generation-completeness assessment identified 45 attempts that did not satisfy the candidate-eligibility criteria: 44 in the single-prompt condition and one in the multi-agent condition. The candidate-level dataset therefore contains 1,455 eligible candidates, comprising 706 single-prompt candidates and 749 multi-agent candidates. The 45 ineligible attempts remain part of generation-completeness reporting.

Each selected Java file is treated as a refactoring target, and each eligible output is treated as a candidate rather than as a directly adoptable patch. Candidates are evaluated in isolated workspaces and are not written back to the source repositories. This design supports controlled before--after measurement but does not evaluate complete project integration.

The dependency-aware single-prompt condition corresponds to B0-V2, and the LangGraph-based multi-agent condition corresponds to P-MA. We use the term \emph{dependency-aware} because both conditions receive repository-level context containing public and protected APIs, caller-used APIs, downstream callers, imports, type dependencies, annotations, inheritance relations, and implemented interfaces.

An eligible candidate is indexed as

\begin{equation}
  c=(p,f,m,o),
  \label{eq:candidate-index}
\end{equation}

where $p$ is the project, $f$ is the selected file, $m$ is the model configuration, and $o$ is the generation condition. RQ1 uses the selected file $(p,f)$ as the unit of analysis. RQ2 and RQ4 use the eligible candidate as the unit of analysis. RQ3 uses eligible matched pairs or complete model blocks, depending on the comparison. RQ5 uses the source file as the primary unit for pre-generation characteristics and the candidate as a secondary sensitivity-analysis unit.

For paired comparisons between generation conditions, the pairing key is

\begin{equation}
  \kappa=(p,f,m).
  \label{eq:pairing-key}
\end{equation}

There are 750 possible file--model pairing keys. A pair is complete and eligible only when both its single-prompt and multi-agent attempts satisfy the candidate-eligibility criteria. This produces 705 complete eligible pairs. The remaining 45 pairing keys are incomplete because one condition did not produce an eligible candidate; no pairing key has ineligible outputs under both conditions. Paired RQ3 analyses use only the 705 complete eligible pairs, and missing partner candidates are not imputed.

For model-level repeated-measures analyses, a complete model block requires eligible observations for all five models for the same file and generation condition. The analysis contains 108 complete five-model blocks for the single-prompt condition and 149 complete five-model blocks for the multi-agent condition. These block-specific denominators are reported with the corresponding statistical tests.

\subsection{Subject Systems and File Selection}
\label{subsec:subject-systems}

We evaluate the candidates on 15 open-source Java repositories selected to provide variation in project size, application domain, and codebase structure. The repositories were required to be primarily implemented in Java, publicly available, non-trivial, suitable for static analysis, and large enough to support repository-level exposure analysis and file-level sampling. They cover GUI frameworks, project-management tools, modelling tools, parsers, build tools, testing frameworks, logging frameworks, mocking frameworks, collections libraries, static-analysis tools, and desktop applications. Table~\ref{tab:subject-systems} summarizes the systems. KLOC denotes physical lines of production Java source code, rounded to thousands; class and method counts provide descriptive indicators of project scale.

\begin{table}[t]
\centering
\caption{Subject systems used in the study.}
\label{tab:subject-systems}
\tiny
\setlength{\tabcolsep}{3pt}
\renewcommand{\arraystretch}{1.08}
\resizebox{\columnwidth}{!}{%
\begin{tabular}{r l r r r r}
\toprule
\rowcolor{gray!15}
\textbf{\#} & \textbf{Project} & \textbf{Java files} & \textbf{Classes} & \textbf{Methods} & \textbf{KLOC} \\
\midrule
1  & Ant                 & 993     & 1{,}365 & 10{,}184 & 239 \\
2  & ArgoUML             & 1{,}687 & 2{,}017 & 11{,}846 & 332 \\
3  & Checkstyle          & 2{,}273 & 2{,}839 & 7{,}215  & 188 \\
4  & Eclipse Collections & 997     & 1{,}360 & 16{,}502 & 214 \\
5  & GanttProject        & 570     & 837     & 4{,}727  & 86 \\
6  & Guava               & 1{,}995 & 4{,}139 & 21{,}018 & 475 \\
7  & JabRef              & 2{,}707 & 3{,}573 & 20{,}262 & 516 \\
8  & jEdit               & 615     & 1{,}143 & 5{,}883  & 193 \\
9  & JHotDraw            & 685     & 773     & 5{,}897  & 122 \\
10 & jsoup               & 97      & 206     & 1{,}250  & 29 \\
11 & JUnit4              & 219     & 246     & 1{,}179  & 20 \\
12 & Logback             & 661     & 660     & 2{,}735  & 59 \\
13 & Mockito             & 521     & 587     & 2{,}138  & 48 \\
14 & SLF4J               & 155     & 151     & 912      & 20 \\
15 & Xerces-J            & 1{,}109 & 1{,}308 & 8{,}004  & 297 \\
\bottomrule
\end{tabular}%
}
\end{table}

We analyze production Java files only. Test files, generated files, build outputs, version-control folders, and files that cannot be processed by the configured profiling pipeline are excluded from the eligible file pool. Test and generated files are removed using path- and filename-based filters, including common test-source, generated-code, and build-output patterns.

We select ten production files from each repository, yielding 150 target files. The sample is intentionally constructed rather than uniformly sampled. A file is eligible when it fits within the generation budget, has at least one code-quality violation detected by the PMD Source Code Analyzer, and exposes at least one repository-facing indicator, such as a public or protected API, caller evidence, an annotation, an inheritance relation, or an implemented interface. Files with no detected fan-in, fan-out, public API, or caller evidence are excluded because the configured profiler exposes no observable repository-facing ripple surface for them.

Selection combines local refactoring-need indicators, including PMD detections and LOC, with repository-exposure indicators, including fan-in, fan-out, public API count, caller-used API count, downstream callers, annotations, inheritance, implemented interfaces, and a heuristic dependency-risk score. The automated selection procedure uses \texttt{DEFAULT\_SEED = 42}. It applies the eligibility criteria and stratifies the files across high-, medium-, and low-fan-in groups; high-public-API files; files combining high smell counts with dependency exposure; and random selection from the remaining eligible files. The multi-agent condition uses the same 150 $(p,f)$ pairs selected for the single-prompt condition and therefore does not resample target files.

This selection supports the investigation of screening behaviour on refactoring-relevant files with observable repository exposure. It does not support prevalence claims about all production Java files or all potential LLM-generated refactorings.

\subsection{Repository Profiling and Dependency Context}
\label{subsec:repository-profiling}

For each repository, we construct a profile using a custom regular-expression-based static-analysis pipeline over Java source text. The study uses this lightweight extractor rather than a type-resolving abstract syntax tree framework such as JavaParser, Spoon, or Eclipse JDT. PMD supplies rule-based Java code-quality detections used as operational indicators of refactoring need~\cite{pmd}; it is not used to extract packages, APIs, callers, or repository dependency graphs.

At file level, the extractor records package declarations, declared classes, interfaces, enums, records, imports, annotations, inheritance relations, implemented interfaces, public and protected methods, constructors, method-call patterns, LOC, and extraction status. Public and protected API signatures are normalized into signature strings, and constructors are recorded separately. Method-call extraction uses textual receiver--method patterns. Consequently, caller-exposure and API-change fields are structural approximations rather than complete semantic or interprocedural analyses.

At repository level, the dependency scanner records dependencies and reverse dependencies between production files. A dependency is recorded when a file imports, references, extends, implements, or calls a type or member declared in another production file. Fan-out is the size of the outgoing dependency set, and fan-in is the number of production files that depend on the selected file. Reverse dependencies are obtained by inverting the file-level dependency graph.

Caller exposure is computed separately. For each selected file, the scanner identifies downstream files that reference the target class and match public or protected API call patterns. Caller-used APIs are public or protected APIs referenced by at least one detected downstream caller. These relationships support RQ1 exposure characterization and the subsequent structural caller-impact check, but they do not constitute a type-resolved call graph.

For each selected file, we construct a dependency-context record containing its path, package, public and protected APIs, caller-used APIs, downstream callers, imports, type dependencies, annotations, inheritance relations, implemented interfaces, fan-in, fan-out, and preservation constraints. The same dependency-context record is supplied for a given file across all models and both generation conditions.

We represent the pre-generation exposure of file $f$ as

\begin{equation}
\begin{aligned}
E(f)=\langle &\mathrm{FI},\mathrm{FO},\mathrm{API},\mathrm{CUA},\mathrm{DC},
\mathrm{ANN},\\
&\mathrm{INH},\mathrm{IFACE},\mathrm{DR},\mathrm{SM},\mathrm{LOC}\rangle,
\end{aligned}
\label{eq:exposure-vector}
\end{equation}

where $\mathrm{FI}$ and $\mathrm{FO}$ denote fan-in and fan-out; $\mathrm{API}$ denotes public and protected APIs; $\mathrm{CUA}$ denotes caller-used APIs; $\mathrm{DC}$ denotes downstream callers; $\mathrm{ANN}$, $\mathrm{INH}$, and $\mathrm{IFACE}$ denote annotation, inheritance, and interface indicators; $\mathrm{DR}$ denotes the dependency-risk score; $\mathrm{SM}$ denotes PMD-detected smells; and $\mathrm{LOC}$ denotes lines of code.

The dependency-risk score is a fixed heuristic used for file selection and exposure analysis:

\begin{equation}
\begin{split}
\mathrm{depRisk}(f)={}&0.40\cdot\mathrm{fan\_in}
+0.35\cdot|\mathrm{CUA}|\\
&+0.15\cdot|\mathrm{API}|
+0.05\cdot\mathbb{I}[\mathrm{inh}]
+0.05\cdot\mathbb{I}[\mathrm{iface}].
\end{split}
\label{eq:dependency-risk-score}
\end{equation}

The weights prioritize inbound exposure and caller-used APIs. The score is a sampling and characterization heuristic, not a validated probability model, and it is not interpreted causally.

The profiler can miss or misclassify relations involving reflection, dynamic dispatch, dependency injection, wildcard imports, name collisions, generated sources, or unusual repository layouts. It may also normalize signatures imperfectly. We therefore describe its outputs as detected structural indicators, report incomplete reason-tag coverage where applicable, and treat extractor precision as a construct-validity limitation.

\subsection{Experimental Setup}
\label{subsec:experimental-setup}

Both generation conditions receive the same selected file, PMD detections, and repository dependency context, but operationalize candidate generation differently.

The single-prompt condition, B0-V2, uses one OpenRouter chat-completion interaction per attempt, followed by the shared verification pipeline. Its prompt contains the complete target file, prioritized smell-specific instructions, dependency context, caller-used APIs, downstream callers, and preservation constraints. The model is instructed to return one complete Java file in a single code block and to preserve package declarations, public and protected APIs, caller-used methods, constructors, annotations, inheritance and interface relations, and externally visible contracts. Private helper methods and internal simplifications are permitted.

The LangGraph-based multi-agent condition, P-MA, uses the same PMD detector, dependency-context record, and ripple-verification implementation. It decomposes processing into the stages summarized in Table~\ref{tab:ma-workflow}.

\begin{table*}[t]
\centering
\caption{Stages of the LangGraph-based multi-agent generation condition.}
\label{tab:ma-workflow}
\begin{tabular}{p{0.22\linewidth}p{0.32\linewidth}p{0.36\linewidth}}
\toprule
\rowcolor{gray!15}
\textbf{Stage group} & \textbf{Nodes} & \textbf{Responsibility} \\
\midrule
Preparation & \textit{load}, \textit{analyze}, \textit{dependency\_context} &
Load the fixed baseline file, detect PMD smells, and construct dependency context. \\
Planning & \textit{plan}, \textit{llm\_plan}, \textit{feasibility} &
Map smells to refactoring techniques, produce a structured plan, and check size or feasibility constraints. \\
Generation & \textit{refactor} &
Generate the candidate Java file. \\
Verification & \textit{verify}, \textit{ripple\_verify}, \textit{llm\_verify} &
Run static quality checks, apply ripple-risk verification, and record an LLM-based assessment. \\
Reporting & \textit{report} &
Persist metrics, traces, and reports. \\
\bottomrule
\end{tabular}
\end{table*}

The multi-agent workflow is not an unbounded repair loop. If generation returns an identical file, the refactoring step may retry with additional instructions up to two times. At study level, a non-terminal multi-agent failure may be retried up to three times for the same file--model pair; a completed record is not overwritten. The condition therefore evaluates staged orchestration rather than repeated repair until a candidate satisfies every screening criterion.

All model interactions use OpenRouter's chat-completions interface~\cite{openrouter}. Table~\ref{tab:model-configurations} reports the configured identifiers. These identifiers denote the routed model configurations used during the study and should not be interpreted as immutable model binaries.

\begin{table}[t]
\centering
\caption{Model configurations used in the study.}
\label{tab:model-configurations}
\begin{tabular}{p{0.22\linewidth}p{0.68\linewidth}}
\toprule
\rowcolor{gray!15}
\textbf{Model} & \textbf{Configured model identifier} \\
\midrule
Gemini & \texttt{google/gemini-3.1-pro-preview} \\
GPT & \texttt{openai/gpt-5.5} \\
Claude & \texttt{anthropic/claude-opus-4.8} \\
DeepSeek & \texttt{deepseek/deepseek-v4-pro} \\
Qwen & \texttt{qwen/qwen3.6-plus} \\
\bottomrule
\end{tabular}
\end{table}

Table~\ref{tab:run-settings} summarizes the principal invocation settings. Target files, model configurations, dependency context, refactoring temperature, and the shared verification implementation are aligned across conditions. Computational cost is not matched because the multi-agent condition adds planning and verification calls.

\begin{table*}[t]
\centering
\caption{Model invocation settings used in the study.}
\label{tab:run-settings}
\begin{tabular}{p{0.22\linewidth}p{0.26\linewidth}p{0.42\linewidth}}
\toprule
\rowcolor{gray!15}
\textbf{Category} & \textbf{Setting} & \textbf{Value} \\
\midrule
\multirow{3}{*}{Temperature}
& Refactoring call & 0.4 for B0-V2 and the MA refactoring agent \\
& MA planning call & 0.2 \\
& MA verification call & 0.1 \\
\midrule
\multirow{2}{*}{Output budget}
& Refactoring output & LOC-scaled, with a default maximum of 2,500 tokens \\
& MA planning / verification & 4,096 / 2,048 tokens \\
\midrule
\multirow{3}{*}{Timeout}
& Refactoring call & LOC-banded, typically 480--3,600 seconds \\
& MA planning / verification & 120 seconds \\
& Candidate / study execution & 600-second candidate HTTP timeout; 3,600-second study timeout \\
\midrule
\multirow{2}{*}{Output format}
& Refactoring output & Complete Java file in one Java code block \\
& MA planning / verification & JSON only \\
\midrule
\multirow{2}{*}{Retry policy}
& Identical-output retry & Up to two retries within the refactoring step \\
& MA pair attempt & Up to three attempts for non-terminal MA failures \\
\bottomrule
\end{tabular}
\end{table*}

\subsection{Generation Completeness and Candidate Evaluation}
\label{subsec:candidate-evaluation}

Each generation attempt produces a persistent candidate-level outcome record. Where retained, the study also stores the source file, generated candidate, prompt or agent trace, model response, diff, before--after profile, and verification report. The tabular records constitute the quantitative outcome data, while the retained artifacts provide additional traceability and support independent inspection where available.

We report analytical eligibility separately from the presence of a generation-attempt record. For condition $o$, let $N_o^{\mathrm{attempted}}$ denote the number of attempted generations and $N_o^{\mathrm{eligible}}$ denote the number that produced a complete candidate suitable for analysis. The candidate-eligibility rate is

\begin{equation}
\mathrm{ER}_o=
\frac{N_o^{\mathrm{eligible}}}
     {N_o^{\mathrm{attempted}}}.
\label{eq:completion-rate}
\end{equation}

The resulting rates are $706/750$ for single-prompt generation and $749/750$ for multi-agent generation. Incomplete or structurally invalid generation outcomes are reported separately and do not inherit contract-risk or adoption-readiness labels derived from partial outputs.

For eligible candidates, local change measures describe the edited file. PMD supplies before and after rule detections. Let $s^{-}$ and $s^{+}$ denote the before and after smell counts. We define

\begin{equation}
\Delta s=s^{+}-s^{-},
\label{eq:smell-delta}
\end{equation}

where a negative value indicates a reduction. Let $\ell^{+}$ and $\ell^{-}$ denote lines added and removed. Churn is defined as

\begin{equation}
\mathrm{churn}=\ell^{+}+\ell^{-}.
\label{eq:churn}
\end{equation}

Where available, we also record methods added or removed, public methods removed, signature changes, import changes, and type-dependency changes. Churn, lines added, and lines removed are available for all 706 eligible single-prompt candidates but for only 72 of the 749 eligible multi-agent candidates. Missing telemetry is retained as missing and is not replaced with zero. Smell delta is available for all eligible candidates.

The verification pipeline organizes evidence into five levels. Level~1 compares the before and after structural profiles and records changes to package declarations, public and protected signatures, constructors, caller-used APIs, imports, type dependencies, annotations, inheritance relations, and implemented interfaces. Level~2 detects structural caller impact by checking whether a previously identified downstream caller references a caller-used API that is absent from the after-profile. Level~3 is reserved for repository- or module-level compilation evidence. Candidate-level compilation was not executed in this experiment, so L3 is reported as unavailable rather than interpreted as zero compilation failures. Level~4 records the behavioural-equivalence indicator derived from detected changes to exception handling, conditional logic, critical calls, and framework-related constructs. Level~5 records targeted Maven test evidence where relevant tests are discoverable and executable.

L1, L2, and L4 fields are available for all 1,455 eligible candidates. L2 produces a broken-caller count, where zero indicates that the configured structural check did not detect a broken caller. A zero count does not establish dynamic caller correctness.

L5 identifies 395 multi-agent candidates as test eligible. Of these, 65 have conclusive test outcomes: 60 preserve the targeted tests and five exhibit regressions. The remaining 330 test-eligible candidates have inconclusive outcomes. No single-prompt candidate has L5 evidence. Therefore, L5 supports a subset analysis but not a uniform comparison between the two generation conditions.

Complete candidate artifact bundles are available for independent inspection for 187 eligible candidates. The remaining candidates are represented by their persistent outcome records but do not have the complete artifact bundles required for independent source-level re-execution. Analyses requiring direct candidate inspection are therefore restricted to the subset for which the necessary artifacts are available.

\subsection{Risk Rules and Screening Outcomes}
\label{subsec:risk-rules}

Contract risk is assigned from the Level~1 profile difference. High-producing conditions include an affected caller-used API, a changed public or protected signature, a removed public API, a changed constructor signature, a changed package declaration, a changed annotation, a changed inheritance or interface declaration, or a large dependency delta.

The dependency delta is defined as

\begin{equation}
\begin{aligned}
\Delta_{\mathrm{dep}}={}&N_{\mathrm{imports\ added}}
+N_{\mathrm{imports\ removed}}\\
&+N_{\mathrm{type\ deps\ added}}
+N_{\mathrm{type\ deps\ removed}}\\
&+|\Delta_{\mathrm{fan\mbox{-}out}}|.
\end{aligned}
\label{eq:dependency-delta}
\end{equation}

The implementation emits the reason tag
\texttt{large\_dependency\_delta:$N$} and assigns High contract risk when

\begin{equation}
\Delta_{\mathrm{dep}}\geq 5.
\end{equation}

Medium contract risk represents internal import or dependency changes without a High-producing condition. No eligible candidate receives a Low contract-risk classification. The threshold of five is a configured screening rule rather than an empirically calibrated failure boundary. We therefore conduct one-sided sensitivity analyses using thresholds of 5, 7, 10, 15, and 20 among the 864 eligible candidates for which the dependency-delta value can be reconstructed from usable risk-reason tags. Thresholds below five cannot be reconstructed because exact dependency-delta values are not retained for candidates that did not trigger the configured threshold rule.

Observed caller-impact risk is derived separately from Level~2. If at least one downstream caller is detected as referring to a caller-used API that is absent from the after-profile, observed caller-impact risk is High; otherwise, it is Low. This outcome provides structural evidence of a potentially broken caller. It does not represent an executed caller failure or confirmed runtime defect.

Let $r^{c}$ denote contract risk and $r^{o}$ denote observed caller-impact risk. The implementation combines these outcomes as

\begin{equation}
r=\max(r^{c},r^{o}),
\label{eq:risk-aggregation}
\end{equation}

with $\textsc{Low}<\textsc{Medium}<\textsc{High}$. The binary High-risk screening indicator is

\begin{equation}
h=\mathbb{I}[r=\textsc{High}].
\label{eq:high-risk}
\end{equation}

Throughout the analysis, $h$ is described as a \emph{pipeline-defined High-risk screening outcome}. Where the High classification originates specifically from Level~1 evidence, we use the term \emph{High contract/dependency risk}. The classification is not interpreted as a confirmed defect, compilation failure, runtime failure, or failure probability. Its empirical precision as an indicator of semantic failure has not been established.

For RQ4, the available evidence is summarized using four reporting categories: \textsc{Low-risk}, \textsc{Review-needed}, \textsc{High Contract/Dependency Risk}, and \textsc{Inconclusive}. The pipeline determines overall risk by combining contract risk and observed caller-impact risk and retaining the higher risk level. Candidates assigned High overall risk are reported as \textsc{High Contract/Dependency Risk} because all High candidates in the eligible cohort also have High contract risk.

\textsc{Low-risk} indicates that a candidate satisfies the pipeline's configured low-risk conditions under the available evidence. It does not establish complete behavioural equivalence. \textsc{Review-needed} indicates that a candidate does not trigger a High-risk classification but lacks sufficient evidence for the Low-risk category. \textsc{High Contract/Dependency Risk} indicates that the pipeline detects substantial contract or dependency-related risk. \textsc{Inconclusive} indicates that the available verification evidence does not support assignment to one of the other reporting categories. These categories summarize evidence produced by the pipeline. They do not establish semantic equivalence, production readiness, or the presence of an actual defect.

Detailed risk-reason tags are usable for 864 eligible candidates, including all 749 multi-agent candidates and 115 of the 706 single-prompt candidates. Analyses of individual risk mechanisms and alternative dependency thresholds are therefore restricted to this subset. Because reason-tag coverage is incomplete and uneven across the two generation conditions, the 115 covered single-prompt candidates are not assumed to represent the entire single-prompt sample.

\subsection{Data Analysis}
\label{subsec:data-analysis}

We use descriptive statistics, paired and repeated-measures comparisons, coverage-aware mechanism summaries, and exploratory association analyses. Counts and proportions are always reported with their applicable denominators. Because many outcomes are binary, skewed, tied, zero-inflated, or ordinal, we use non-parametric tests where appropriate and report effect sizes alongside $p$-values. Holm adjustment controls multiple comparisons within each declared test family~\cite{holm1979simple}.

\textbf{RQ1: repository exposure.}
The unit of analysis is the selected file $(p,f)$, with $N=150$. We summarize LOC, PMD detections, fan-in, fan-out, public and protected API count, caller-used API count, downstream callers, dependency-risk score, inheritance, and interface implementation using counts, prevalence, medians, interquartile ranges, means, minima, and maxima. Project-level distributions are compared using Kruskal--Wallis tests~\cite{kruskal1952use}, with Holm adjustment across the exposure measures. Spearman correlations~\cite{spearman1904proof} describe relationships among file-local and repository-exposure indicators.

\textbf{RQ2: detected risk indicators.}
The unit of analysis is the eligible candidate $(p,f,m,o)$, with $N=1{,}455$. We report contract-risk levels, the pipeline-defined High-risk screening outcome, detected broken-caller occurrence and counts, the behavioural-equivalence indicator, and verification coverage. Counts and percentages for fine-grained risk reasons use the reason-covered denominator of 864 rather than the complete eligible dataset. Wilson intervals~\cite{wilson1927probable} are used for selected binary proportions. Contract risk, broken-caller evidence, the behavioural-equivalence indicator, test evidence, compilation availability, and artifact availability are reported as separate constructs.

\textbf{RQ3: generation-condition and model comparisons.}
Generation-condition comparisons use the 705 complete eligible pairing keys $\kappa=(p,f,m)$. Exact two-sided McNemar tests~\cite{mcnemar1947note} compare paired binary outcomes: the High-risk screening outcome, broken-caller occurrence, and a false behavioural-equivalence indicator. The paired difference in broken-caller counts is analyzed using the Wilcoxon signed-rank test~\cite{wilcoxon1945individual}. Because the distribution is strongly zero-inflated, we report the number of non-zero paired differences and identify the normal approximation used.

Fine-grained risk-reason flags are compared only among the 115 pairs for which both candidates have usable reason tags. Exact McNemar tests are applied to 11 reason flags, with Holm adjustment across the 11-test family.

Model-level comparisons use complete five-model blocks separately for each generation condition: 108 single-prompt blocks and 149 multi-agent blocks. Cochran's $Q$ test~\cite{cochran1950comparison} is used for matched binary outcomes, and the Friedman test~\cite{friedman1937use} is used for broken-caller counts. Pairwise exact McNemar comparisons with Holm adjustment are reported where the corresponding Cochran test is significant. When a post-hoc comparison is unavailable, we report the significant omnibus result without claiming which individual model pairs differ.

\textbf{RQ4: staged screening and verification evidence.}
RQ4 uses all 1,455 eligible candidates and examines their distribution across the four adoption-readiness categories described above. We report the evidence available at each verification level while keeping structural risk, detected caller impact, the behavioural-equivalence indicator, compilation availability, targeted-test outcomes, and artifact availability analytically distinct.

Test findings are reported relative to the 395 test-eligible candidates and, separately, the 65 candidates with conclusive test outcomes. Of the conclusive outcomes, 60 preserve the targeted tests and five exhibit regressions. Because candidate-level compilation was not performed, we do not infer a compilation-success or compilation-failure rate. Differences across generation conditions and model configurations are analyzed under RQ3 rather than repeated as independent RQ4 tests.

\textbf{RQ5: exploratory associated characteristics.}
RQ5 analyzes three outcomes separately: pipeline-defined High contract/dependency risk, detected broken-caller occurrence, and a false behavioural-equivalence indicator. For pre-generation source-file characteristics, the primary unit of analysis is the unique source file $(p,f)$, with $N=150$. Each file's eligible candidates are aggregated into the proportion exhibiting each outcome. Spearman correlations relate continuous file characteristics to these proportions. Binary file characteristics are compared using Mann--Whitney tests on the file-level proportions, with Cliff's $\delta$ as the effect size. Holm adjustment is applied separately to the ten pre-generation factors for each of the three outcome families.

Candidate-level analyses of the same factors are retained as sensitivity analyses. Numeric factors are compared using Mann--Whitney tests with Cliff's $\delta$, while binary factors are assessed using two-sided Fisher exact tests and odds ratios. These analyses are secondary because candidates generated from the same source file and repository are not statistically independent. Model- and condition-specific rates are reported descriptively and interpreted alongside the RQ3 findings rather than tested again as independent RQ5 results.

Post-generation change measures are analyzed separately as concurrent indicators rather than antecedent causes. Available-case Mann--Whitney analyses compare churn, lines added, lines removed, and smell delta for each outcome, with Holm adjustment within each outcome family. Missing churn and line-change values are not imputed. Risk-reason flags are summarized as mechanisms contributing to the High classification. Because several of these flags directly define the outcome, they are not tested or interpreted as independent explanatory predictors.

Across RQ5, file-level aggregation reduces repeated-candidate dependence but does not model residual dependence among files drawn from the same repository. The analyses are exploratory and support hypotheses about screening and verification priorities rather than causal or predictive claims.

%% file: sections/04-results-and-analysis.tex
\section{Results and Analysis}
\label{sec:results}

This section reports the empirical results for the five research questions.
RQ1 first characterizes the repository exposure of the selected Java files before generation.
RQ2 then reports the repository-level ripple-risk indicators detected in generated candidates.
RQ3 compares how risk profiles differ across generation conditions and LLM configurations.
RQ4 maps the available verification evidence into adoption-readiness classes.
RQ5 examines which source-file characteristics and generated-change indicators are associated with High contract and dependency risk, detected broken-caller evidence, and a false behavioural-equivalence indicator.

\subsection{Repository Exposure (RQ1)}
\label{subsec:rq1-results}

RQ1 examines the repository exposure of the 150 Java files selected for refactoring. The unit of analysis is one production Java file. Each of the 15 subject systems contributes ten files, and complete measurements are available for all reported RQ1 indicators. Because file selection required both evidence of refactoring need and an observable repository-facing surface, the resulting distributions characterize the selected cohort rather than Java files in general.

We examine two complementary dimensions. First, LOC and baseline PMD detections characterize the size and local refactoring need of the selected files. Second, fan-in, fan-out, public and protected APIs, caller-used APIs, downstream callers, inheritance relations, implemented interfaces, and the dependency-risk score characterize their exposure to the surrounding repository.

\begin{table*}[!t]
\centering
\scriptsize
\setlength{\tabcolsep}{2.5pt}
\caption{Project-level characteristics of the selected files before LLM-based refactoring.}
\label{tab:rq1-project-exposure}
\resizebox{\textwidth}{!}{%
\begin{tabular}{l c cc cc cc cc cc cc cc cc cc}
\toprule
\rowcolor{gray!15}
\multirow{2}{*}{\textbf{Project}}
& \multirow{2}{*}{\textbf{Files}}
& \multicolumn{2}{c}{\textbf{LOC}}
& \multicolumn{2}{c}{\textbf{Smells}}
& \multicolumn{2}{c}{\textbf{Fan-in}}
& \multicolumn{2}{c}{\textbf{Fan-out}}
& \multicolumn{2}{c}{\textbf{APIs}}
& \multicolumn{2}{c}{\textbf{Caller APIs}}
& \multicolumn{2}{c}{\textbf{Callers}}
& \multicolumn{2}{c}{\textbf{Dep. risk}}
& \multicolumn{2}{c}{\textbf{Structural}} \\
\cmidrule(lr){3-4}
\cmidrule(lr){5-6}
\cmidrule(lr){7-8}
\cmidrule(lr){9-10}
\cmidrule(lr){11-12}
\cmidrule(lr){13-14}
\cmidrule(lr){15-16}
\cmidrule(lr){17-18}
\cmidrule(lr){19-20}
\rowcolor{gray!15}
&
& \textbf{Tot.} & \textbf{Med.}
& \textbf{Tot.} & \textbf{Med.}
& \textbf{Tot.} & \textbf{Med.}
& \textbf{Tot.} & \textbf{Med.}
& \textbf{Tot.} & \textbf{Med.}
& \textbf{Tot.} & \textbf{Med.}
& \textbf{Tot.} & \textbf{Med.}
& \textbf{Tot.} & \textbf{Med.}
& \textbf{Inh.} & \textbf{Iface} \\
\midrule

Ant
& 10 & 3258 & 254.0 & 48 & 3.0
& 108 & 1.5 & 156 & 15.0
& 145 & 10.5 & 44 & 1.0
& 108 & 1.5 & 37.7 & 2.65 & 7 & 4 \\

ArgoUML
& 10 & 4709 & 460.5 & 53 & 5.0
& \textbf{206} & \textbf{0.0} & 325 & 16.0
& 89 & 4.5 & 24 & 0.0
& \textbf{206} & \textbf{0.0} & 22.3 & 0.98 & 8 & 4 \\

Checkstyle
& 10 & 3849 & 369.0 & 12 & 1.0
& 93 & 1.5 & 160 & 17.0
& 137 & 5.5 & 12 & 1.0
& 93 & 1.5 & 25.2 & 1.05 & 7 & 2 \\

Eclipse Collections
& 10 & 5793 & 261.5 & 95 & 2.5
& 44 & 1.0 & \textbf{345} & \textbf{29.5}
& \textbf{383} & \textbf{17.0}
& \textbf{198} & 2.0
& 44 & 1.0 & \textbf{127.0} & 3.30 & 0 & 5 \\

GanttProject
& 10 & 2353 & 200.0 & 41 & 2.0
& 91 & 2.0 & 148 & 7.5
& 119 & 7.5 & 59 & 1.5
& 91 & 2.0 & 38.9 & 1.73 & 4 & 4 \\

Guava
& 10 & 3470 & 221.0 & 72 & 4.0
& 87 & 2.5 & 236 & 19.5
& 136 & 11.5 & 36 & 1.0
& 87 & 2.5 & 33.3 & 2.47 & 2 & 3 \\

JabRef
& 10 & 1500 & 124.5 & 29 & 1.0
& 14 & 1.0 & 174 & 17.5
& 49 & 3.0 & 17 & 1.0
& 14 & 1.0 & 13.7 & 0.73 & 2 & 5 \\

jEdit
& 10 & 2466 & 232.5 & 105 & 12.5
& 107 & 1.0 & 109 & 7.5
& 54 & 5.0 & 36 & 1.5
& 107 & 1.0 & 21.3 & 1.43 & 6 & 5 \\

JHotDraw
& 10 & 3396 & 163.5 & 122 & 6.0
& 29 & 3.0 & 187 & 6.5
& 163 & 6.0 & 40 & 2.5
& 29 & 3.0 & 39.0 & 2.32 & 7 & 4 \\

jsoup
& 10 & 4196 & 297.0 & 177 & 13.5
& 43 & 3.0 & 160 & 10.0
& 182 & 6.5 & 98 & 5.0
& 43 & 3.0 & 61.9 & 2.60 & 5 & 2 \\

JUnit4
& 10 & 1627 & 104.0 & 57 & 4.0
& 166 & 4.0 & 90 & 5.5
& 112 & 5.5 & 66 & 2.0
& 166 & 4.0 & 40.4 & 1.18 & 3 & 7 \\

Logback
& 10 & \textbf{1127} & \textbf{98.5} & 53 & 3.0
& 3 & 0.0 & 66 & 5.0
& 39 & 3.0 & 6 & 0.0
& 3 & 0.0 & 8.3 & 0.48 & 5 & 1 \\

Mockito
& 10 & 2652 & 168.5 & 50 & 2.0
& 34 & 1.0 & 135 & 7.5
& 229 & 10.0 & 21 & 1.5
& 34 & 1.0 & 42.1 & 2.90 & 1 & 6 \\

SLF4J
& 10 & 2624 & 236.0 & 117 & 8.5
& 27 & 1.0 & 80 & 8.0
& 147 & 8.5 & 112 & 4.5
& 27 & 1.0 & 61.7 & 2.45 & 1 & 7 \\

Xerces-J
& 10 & \textbf{8379} & 276.0 & \textbf{269} & \textbf{20.0}
& 194 & \textbf{14.5} & 132 & 8.0
& 186 & 13.5 & 80 & \textbf{7.5}
& 182 & \textbf{14.0} & 56.6 & \textbf{6.15} & 8 & 6 \\

\bottomrule
\end{tabular}%
}

\vspace{0.5em}
\footnotesize
\textit{Note.} Each project contributes ten selected production Java files.
For each metric, \textbf{Tot.} reports the total across the ten selected files, whereas \textbf{Med.} reports the median file-level value.
The totals describe the selected-file cohort and are not full-repository totals.
LOC = lines of code; Smells = PMD-detected code-quality violations; Fan-in = repository files referencing the selected file; Fan-out = repository files on which the selected file depends; APIs = public and protected APIs; Caller APIs = APIs used by detected downstream callers; Callers = detected downstream caller files; Dep. risk = dependency-risk score; Inh. = files with inheritance relations; and Iface = files implementing interfaces.
Bold values identify the principal descriptive contrasts discussed in the text and do not indicate statistical significance.
\end{table*}

Table~\ref{tab:rq1-project-exposure} shows substantial descriptive variation across the projects despite the equal allocation of ten files per system. The amount of selected production code ranges from 1,127 LOC in Logback to 8,379 LOC in Xerces-J. Baseline PMD detections range from 12 in Checkstyle to 269 in Xerces-J. Thus, equal numbers of selected files do not imply equal code volume or equal baseline refactoring need.

Repository-exposure profiles also differ across the selected project samples. ArgoUML has 206 total inbound references and 206 detected downstream callers, but a median of zero for both indicators. Its inbound exposure is therefore concentrated in a small number of selected files. By contrast, Xerces-J has a median fan-in of 14.5 and a median of 14 downstream callers, indicating that inbound exposure is distributed more broadly across its selected files. Eclipse Collections has the largest aggregate outbound and API surfaces, with 345 fan-out relations, 383 public and protected APIs, and 198 caller-used APIs. These contrasts demonstrate why project totals alone are insufficient and should be interpreted together with the corresponding file-level medians.

\begin{table}[!t]
\centering
\scriptsize
\setlength{\tabcolsep}{2.5pt}
\caption{Distribution of exposure indicators across the 150 selected files.}
\label{tab:rq1-overall-exposure}
\resizebox{0.88\columnwidth}{!}{%
\begin{tabular}{lrrrrrrr}
\toprule
\rowcolor{gray!15}
\textbf{Metric}
& \textbf{Min}
& \textbf{Q1}
& \textbf{Med.}
& \textbf{Mean}
& \textbf{Q3}
& \textbf{Max}
& \textbf{\% $>0$} \\
\midrule

LOC
& 18 & 88.5 & 212.5 & 342.7 & 375.5 & 3182 & 100.0 \\

Smells
& 1 & 1.0 & 3.0 & 8.7 & 11.8 & 60 & 100.0 \\

Fan-in
& 0 & 0.0 & 1.0 & 8.3 & 5.8 & 181
& \textbf{68.7} \\

Fan-out
& 0 & 4.0 & 10.0 & 16.7 & 21.8 & 119
& \textbf{98.0} \\

APIs
& 0 & 2.0 & 7.0 & 14.5 & 18.0 & 135
& \textbf{92.7} \\

Caller APIs
& 0 & 0.0 & 1.5 & 5.7 & 6.0 & 86
& \textbf{68.7} \\

Callers
& 0 & 0.0 & 1.0 & 8.2 & 5.8 & 181
& \textbf{68.7} \\

Dep. risk
& 0.0 & 0.7 & 1.7 & 4.2 & 4.8 & 43.4
& \textbf{95.3} \\

Inheritance
& 0 & 0.0 & 0.0 & 0.44 & 1.0 & 1
& \textbf{44.0} \\

Interface
& 0 & 0.0 & 0.0 & 0.43 & 1.0 & 1
& \textbf{43.3} \\

\bottomrule
\end{tabular}%
}

\vspace{0.4em}
\scriptsize
\textit{Note.} Values are calculated at file level across all 150 selected files.
The final column reports the percentage of files with a value greater than zero.
LOC = lines of code; Smells = PMD-detected code-quality violations; APIs = public and protected APIs; Caller APIs = APIs used by detected downstream callers; Callers = detected downstream caller files; and Dep. risk = dependency-risk score.
Inheritance and Interface are binary indicators.
Bold values identify the prevalence estimates discussed in the text.
\end{table}

Across the selected cohort, the 150 files contain 51,399 lines of production Java code and 1,300 baseline PMD detections. As shown in Table~\ref{tab:rq1-overall-exposure}, outbound exposure is widespread: 147 files (98.0\%) depend on at least one other production file in the corresponding repository. Similarly, 139 files (92.7\%) expose at least one public or protected API, while 143 files (95.3\%) have a nonzero dependency-risk score.

Inbound exposure is less prevalent but remains observable in a substantial majority of the selected files. A total of 103 files (68.7\%) have nonzero fan-in. The same number have at least one caller-used API and at least one detected downstream caller. Structural relationships are also present in a considerable portion of the cohort: 66 files (44.0\%) participate in an inheritance relation, and 65 files (43.3\%) implement at least one interface.

The distributions reveal considerable concentration in the exposure measures. Fan-in has a median of 1.0 but a mean of 8.3 and a maximum of 181. The number of downstream callers follows a similar pattern, with a median of 1.0, a mean of 8.2, and a maximum of 181. Public and protected API count has a median of seven but reaches a maximum of 135. These differences between medians, means, and maxima show that the cohort contains many files with modest exposure and a smaller number of highly connected files with substantially larger repository-facing surfaces.

\begin{table}[!t]
\centering
\scriptsize
\setlength{\tabcolsep}{4pt}
\caption{Differences among projects in the file-level exposure indicators.}
\label{tab:rq1-project-tests}
\resizebox{\columnwidth}{!}{%
\begin{tabular}{lrrr}
\toprule
\rowcolor{gray!15}
\textbf{Metric}
& \textbf{$H$}
& \textbf{Raw $p$}
& \textbf{Holm-adjusted $p$} \\
\midrule

LOC
& 23.50 & 0.0526 & 0.1863 \\

Baseline smells
& \textbf{41.91} & \textbf{0.0001}
& \textbf{0.0010} \\

Fan-in
& 25.21 & 0.0325 & 0.1863 \\

Fan-out
& 28.56 & 0.0120 & 0.0839 \\

Public/protected APIs
& 15.30 & 0.3578 & 0.3578 \\

Caller-used APIs
& 25.40 & 0.0308 & 0.1863 \\

Downstream callers
& 25.11 & 0.0335 & 0.1863 \\

Dependency-risk score
& 21.34 & 0.0932 & 0.1863 \\

\bottomrule
\end{tabular}%
}

\vspace{0.4em}
\scriptsize
\textit{Note.} Kruskal--Wallis tests compare the file-level distributions across the 15 projects, with ten selected files per project and $df=14$. Holm adjustment is applied across the eight comparisons. Bold indicates the only result that remains statistically significant at $\alpha=0.05$ after adjustment.
\end{table}

Table~\ref{tab:rq1-project-tests} compares the file-level distributions across projects. Although fan-in, fan-out, caller-used APIs, and downstream callers produce unadjusted $p$-values below 0.05, none remains statistically significant after adjustment for the eight comparisons. Baseline PMD detections constitute the only statistically significant project difference after Holm correction ($H=41.91$, adjusted $p=0.001$). The observed project-level contrasts in the repository-exposure measures are therefore treated as descriptive heterogeneity rather than evidence of statistically established differences among the projects.

Local file characteristics provide only limited information about inbound repository exposure. LOC is weakly correlated with fan-in ($\rho=0.22$) and the number of downstream callers ($\rho=0.22$). Baseline PMD detections have still weaker correlations with fan-in ($\rho=0.14$) and downstream callers ($\rho=0.14$). File size and baseline code-quality detections therefore do not adequately capture how strongly a file is connected to the surrounding repository. This distinction supports analyzing repository exposure explicitly rather than inferring it from local file characteristics.

\begin{rqanswer}
\noindent\faIcon{lightbulb}~\textbf{Answer to RQ1:}
The selected files exhibit substantial but uneven repository exposure. Almost all have outbound dependencies, and more than 90\% expose at least one public or protected API. Approximately two thirds have inbound references, caller-used APIs, and detected downstream callers. The exposure distributions are strongly right-skewed: most selected files have modest inbound exposure, whereas a smaller group has considerably larger repository-facing surfaces. After correcting for multiple comparisons, the exposure indicators do not differ significantly across projects, although baseline PMD detections do. Moreover, file size and baseline PMD detections are only weakly associated with inbound exposure. Repository-level indicators therefore provide information that cannot be inferred reliably from local file characteristics alone.
\end{rqanswer}

\subsection{Detected Ripple-Risk Indicators (RQ2)}
\label{subsec:rq2-results}

RQ2 examines the repository-level ripple-risk indicators detected in the eligible refactoring candidates. The unit of analysis is one candidate, identified by project, file, model configuration, and generation condition. Of the 1,500 planned generation attempts, 45 were confirmed as incomplete or otherwise invalid generations and were excluded from candidate-level analysis. The resulting RQ2 cohort contains 1,455 eligible candidates: 706 generated using the single-prompt condition and 749 generated using the multi-agent condition.

We examine three distinct forms of evidence: contract and dependency risk, detected downstream caller impact, and behavioural-risk indicators. These measures capture different properties of a candidate and are therefore reported separately. In particular, the pipeline-defined High-risk classification represents the activation of one or more structural screening rules. It is not interpreted as a confirmed defect, compilation failure, runtime failure, or estimate of the probability that a candidate is unsafe. Formal paired comparisons between the generation conditions and model configurations are presented in RQ3.

\begin{table*}[t]
\centering
\scriptsize
\setlength{\tabcolsep}{3.0pt}
\caption{Core repository-level risk indicators detected in the eligible candidates.}
\label{tab:rq2-core-risk}
\resizebox{0.90\textwidth}{!}{%
\begin{tabular}{lrrrrrr}
\toprule
\rowcolor{gray!15}
\textbf{Detected outcome}
& \textbf{Count}
& \textbf{$N$}
& \textbf{Rate}
& \textbf{Wilson 95\% CI}
& \textbf{Single-prompt}
& \textbf{Multi-agent} \\
\midrule

High contract/dependency risk
& 1,365 & 1,455 & 93.8\% & 92.5--94.9\%
& 661/706 (93.6\%)
& 704/749 (94.0\%) \\

Medium contract/dependency risk
& 90 & 1,455 & 6.2\% & 5.1--7.5\%
& 45/706 (6.4\%)
& 45/749 (6.0\%) \\

At least one broken caller
& 106 & 1,455 & 7.3\% & 6.1--8.7\%
& 57/706 (8.1\%)
& 49/749 (6.5\%) \\

Behavioural-equivalence indicator false
& 290 & 1,455 & 19.9\% & 18.0--22.1\%
& 138/706 (19.5\%)
& 152/749 (20.3\%) \\

\bottomrule
\end{tabular}%
}

\vspace{0.5em}
\footnotesize
\textit{Note.} Rates are calculated over the 1,455 eligible candidates. High and Medium refer to the pipeline-defined contract/dependency-risk classification. A broken caller is recorded when the structural analysis detects at least one downstream caller that refers to a caller-used API absent from the generated candidate. The behavioural-equivalence indicator is based on the configured static behavioural checks. These outcomes are screening indicators and do not independently establish semantic failure. Generation-condition values are descriptive; paired comparisons are reported in RQ3.
\end{table*}

As shown in Table~\ref{tab:rq2-core-risk}, 1,365 of the 1,455 eligible candidates (93.8\%) trigger at least one condition associated with High contract/dependency risk. The remaining 90 candidates (6.2\%) are classified as Medium risk, and no eligible candidate receives a Low contract-risk classification. The high prevalence of this outcome reflects the sensitivity of the configured screening rules to changes in caller-used APIs, public or protected signatures, constructors, inheritance and interface declarations, annotations, packages, and repository dependencies. It should therefore be understood as the behaviour of the screening rule rather than as an estimate of the proportion of defective refactorings.

Directly detected caller impact is considerably less frequent. The analysis identifies at least one potentially broken downstream caller for 106 candidates (7.3\%), whereas 1,349 candidates (92.7\%) have no detected broken caller. This does not establish that the latter candidates preserve all callers, because the measure is limited to callers and API relationships recognized by the configured static extractor. Nevertheless, it distinguishes direct caller-impact evidence from the broader contract/dependency-risk classification.

The behavioural-equivalence indicator is false for 290 candidates (19.9\%). This indicator captures changes to the structural patterns used by the pipeline to approximate behavioural risk, including exception handling, conditional logic, critical calls, and framework-related constructs. It does not constitute an execution-based test of semantic equivalence. Contract/dependency risk, broken-caller evidence, and behavioural-risk evidence are consequently retained as separate outcomes throughout the analysis.

Detailed risk reasons are available for 864 of the 1,455 eligible candidates. This subset contains all 749 multi-agent candidates but only 115 of the 706 single-prompt candidates. The following mechanism-level analysis is therefore restricted to these 864 candidates. Because detailed-reason coverage differs substantially between the two conditions, the condition-specific values in Table~\ref{tab:rq2-risk-mechanisms} are descriptive and should not be interpreted as representative comparisons.

\begin{table*}[t]
\centering
\scriptsize
\setlength{\tabcolsep}{3pt}
\caption{Detailed risk mechanisms among candidates with usable risk-reason information.}
\label{tab:rq2-risk-mechanisms}
\resizebox{0.90\textwidth}{!}{%
\begin{tabular}{llrrrr}
\toprule
\rowcolor{gray!15}
\textbf{Category}
& \textbf{Detected mechanism}
& \textbf{Count}
& \textbf{Rate}
& \textbf{Single-prompt}
& \textbf{Multi-agent} \\
\midrule

\multirow{3}{*}{Dependency}
& Any dependency change
& 716 & 82.9\%
& 94/115 (81.7\%)
& 622/749 (83.0\%) \\

& Large dependency delta
& 662 & 76.6\%
& 85/115 (73.9\%)
& 577/749 (77.0\%) \\

& Internal import/dependency change
& 54 & 6.3\%
& 9/115 (7.8\%)
& 45/749 (6.0\%) \\

\midrule

Caller/API
& Caller-used API changed or removed
& 597 & 69.1\%
& 83/115 (72.2\%)
& 514/749 (68.6\%) \\

\midrule

\multirow{5}{*}{Contract/API}
& Any contract/API change
& 333 & 38.5\%
& 40/115 (34.8\%)
& 293/749 (39.1\%) \\

& Public/protected signature changed
& 231 & 26.7\%
& 27/115 (23.5\%)
& 204/749 (27.2\%) \\

& Constructor signature changed
& 138 & 16.0\%
& 19/115 (16.5\%)
& 119/749 (15.9\%) \\

& Public API removed
& 82 & 9.5\%
& 6/115 (5.2\%)
& 76/749 (10.1\%) \\

& Package declaration changed
& 0 & 0.0\%
& 0/115 (0.0\%)
& 0/749 (0.0\%) \\

\midrule

\multirow{2}{*}{Framework/structure}
& Annotation-related change
& 78 & 9.0\%
& 3/115 (2.6\%)
& 75/749 (10.0\%) \\

& Inheritance/interface change
& 28 & 3.2\%
& 3/115 (2.6\%)
& 25/749 (3.3\%) \\

\bottomrule
\end{tabular}%
}

\vspace{0.5em}
\footnotesize
\textit{Note.} The denominator is the 864 eligible candidates with usable detailed risk reasons: 115 single-prompt and 749 multi-agent candidates. Mechanisms are not mutually exclusive, and one candidate may contribute to several rows. Condition-specific rates are descriptive because detailed-reason coverage is incomplete for the single-prompt condition. These mechanisms are components of the configured risk rules and are not independent evidence of actual candidate failure.
\end{table*}

Table~\ref{tab:rq2-risk-mechanisms} shows that dependency-related changes dominate the detailed risk evidence. Among the 864 candidates with usable reasons, 716 (82.9\%) contain at least one dependency change, and 662 (76.6\%) reach the configured large-dependency-delta threshold. An additional 54 candidates (6.3\%) contain an internal import or dependency change without reaching that threshold.

Changes affecting caller-used APIs are also common. The profiler identifies a changed or removed caller-used API in 597 candidates (69.1\%). Because this measure is based on normalized signatures extracted by the static profiler, it indicates a potential contract mismatch rather than confirmed caller failure. The directly detected broken-caller outcome in Table~\ref{tab:rq2-core-risk} provides a narrower form of caller-impact evidence.

Other contract-related changes occur less frequently. Overall, 333 candidates (38.5\%) contain at least one contract/API-change reason. Public or protected signature changes occur in 231 candidates (26.7\%), constructor-signature changes in 138 (16.0\%), and public-API removal in 82 (9.5\%). No package-declaration change is recorded in the reason-covered cohort. Annotation-related changes occur in 78 candidates (9.0\%), while inheritance or implemented-interface changes occur in 28 candidates (3.2\%).

The frequency of large dependency deltas raises the question of whether the High-risk prevalence is an artifact of the configured threshold of five. We therefore recalculate the classification among the 864 candidates for which the numeric dependency delta can be reconstructed. The other High-producing risk conditions are held unchanged while the dependency-delta threshold is increased from 5 to 7, 10, 15, and 20. Thresholds below five cannot be reconstructed reliably from the historical reason data and are not considered.

\begin{table}[t]
\centering
\scriptsize
\setlength{\tabcolsep}{5pt}
\caption{Sensitivity of the High-risk classification to the dependency-delta threshold.}
\label{tab:rq2-threshold-sensitivity}
\begin{tabular}{rrrr}
\toprule
\rowcolor{gray!15}
\textbf{Threshold}
& \textbf{High}
& \textbf{$N$}
& \textbf{Rate} \\
\midrule
5  & 810 & 864 & 93.8\% \\
7  & 790 & 864 & 91.4\% \\
10 & 765 & 864 & 88.5\% \\
15 & 745 & 864 & 86.2\% \\
20 & 714 & 864 & 82.6\% \\
\bottomrule
\end{tabular}

\vspace{0.4em}
\scriptsize
\textit{Note.} The analysis includes only candidates with usable detailed risk reasons. At each threshold, all other High-producing conditions remain unchanged. The analysis is one-sided because thresholds below five cannot be reconstructed from the available historical reason data.
\end{table}

Table~\ref{tab:rq2-threshold-sensitivity} shows that the High-risk rate declines as the dependency-delta threshold becomes more conservative, from 93.8\% at the implemented threshold of five to 82.6\% at a threshold of twenty. The classification is therefore sensitive to the selected threshold, but the large majority of reason-covered candidates remain classified as High even at the most conservative threshold examined. This persistence occurs because candidates can also trigger the classification through caller-used API changes, signature changes, public-API removal, constructor changes, and other contract-related conditions. The sensitivity analysis consequently confirms that the threshold contributes materially to the prevalence of High classifications but does not account for the result on its own.

\begin{figure}[t]
\centering
\includegraphics[width=\columnwidth]{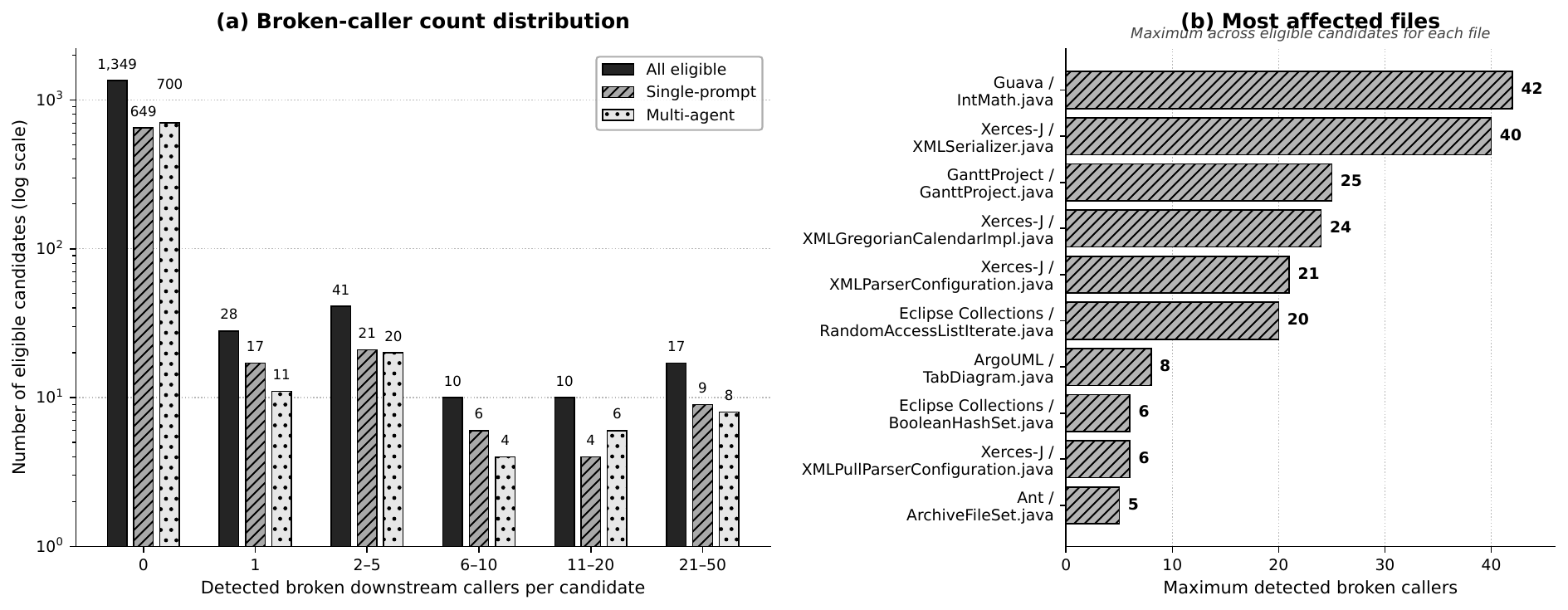}
\caption{Distribution and concentration of detected downstream caller impact among the 1,455 eligible candidates. Panel (a) presents the distribution of broken-caller counts overall and by generation condition. Panel (b) presents the files associated with the largest eligible broken-caller counts. The distribution is strongly zero-inflated, with the largest effects concentrated in a small number of files.}
\label{fig:rq2-broken-caller-distribution}
\end{figure}

Figure~\ref{fig:rq2-broken-caller-distribution} presents the distribution of detected broken-caller counts. Most eligible candidates have no detected caller impact: 1,349 of 1,455 candidates (92.7\%) have a broken-caller count of zero. Among the 106 candidates with a positive count, 28 affect one caller, 41 affect between two and five callers, 10 affect between six and ten callers, 10 affect between 11 and 20 callers, and 17 affect between 21 and 50 callers. No eligible candidate affects more than 50 detected callers.

The largest eligible count is 42, observed for a multi-agent candidate generated for \texttt{IntMath.java} in Guava. Other large counts include 40 for \texttt{XMLSerializer.java} in Xerces-J and 25 for \texttt{GanttProject.java} in GanttProject. Several of the highest-count records could not be independently re-audited because their candidate artifacts were unavailable. These cases are therefore retained as historical pipeline outcomes but should not be interpreted as independently confirmed compilation or runtime failures.

The descriptive condition-level counts show 57 single-prompt and 49 multi-agent candidates with at least one detected broken caller. These unpaired totals do not account for the shared project--file--model structure or the 45 incomplete generation pairs. The formal comparison between conditions is consequently based on the 705 complete eligible pairs and is reported in RQ3.

\begin{rqanswer}
\noindent\faIcon{lightbulb}~\textbf{Answer to RQ2:}
The screening pipeline identifies repository-facing changes in a large majority of the eligible refactoring candidates. Specifically, 93.8\% trigger at least one High contract/dependency-risk condition. Among candidates with detailed reason information, the most frequent signals are dependency changes, large dependency deltas, and changes to caller-used APIs. However, these rule-based outcomes should not be interpreted as the proportion of defective or empirically unsafe refactorings. More direct evidence is considerably less frequent: 7.3\% of eligible candidates have at least one detected broken caller, and the behavioural-equivalence indicator is false for 19.9\%. Raising the dependency-delta threshold reduces the High-risk rate, although most reason-covered candidates remain High because other contract-related conditions can independently trigger the classification.
\end{rqanswer}

\subsection{Generation-Condition and Model Differences (RQ3)}
\label{subsec:rq3-results}

RQ3 examines whether the detected repository-level risk indicators differ across the two generation conditions and five LLM configurations. We first compare single-prompt and multi-agent candidates generated for the same project, file, and model configuration. We then examine whether the detected indicators differ across model configurations within each generation condition.

Of the 750 possible single-prompt and multi-agent pairing keys, 705 contain eligible candidates from both conditions. The remaining 45 pairs are incomplete because 44 single-prompt attempts and one multi-agent attempt did not produce candidates eligible for candidate-level analysis. All condition-level inferential comparisons therefore use the 705 complete pairs. High contract/dependency risk, detected caller impact, the behavioural-equivalence indicator, and broken-caller count are analyzed separately because they represent different aspects of repository-level risk.

\begin{table*}[!t]
\centering
\scriptsize
\setlength{\tabcolsep}{3.2pt}
\caption{Paired comparison of single-prompt and multi-agent generation.}
\label{tab:rq3-generation-condition-comparison}
\resizebox{0.90\textwidth}{!}{%
\begin{tabular}{lrrrrrrl}
\toprule
\rowcolor{gray!15}
\textbf{Outcome}
& \textbf{Paired $N$}
& \textbf{Single-prompt}
& \textbf{Multi-agent}
& \textbf{MA--SP}
& \textbf{Discordant pairs}
& \textbf{$p$}
& \textbf{Test} \\
\midrule

High contract/dependency risk
& 705
& 660 (93.6\%)
& 660 (93.6\%)
& 0.0 pp
& 2 / 2
& 1.000
& Exact McNemar \\

Broken caller detected
& 705
& 57 (8.1\%)
& 49 (7.0\%)
& $-1.1$ pp
& 23 / 15
& 0.256
& Exact McNemar \\

Behavioural-equivalence indicator false
& 705
& 138 (19.6\%)
& 152 (21.6\%)
& $+2.0$ pp
& 55 / 69
& 0.243
& Exact McNemar \\

Broken-caller count
& 705
& 0 [0, 0]
& 0 [0, 0]
& 0
& 47 non-zero differences
& 1.000
& Wilcoxon signed-rank \\

\bottomrule
\end{tabular}%
}

\vspace{0.5em}
\footnotesize
\textit{Note.} Binary outcomes are reported as counts and percentages. For the McNemar tests, discordant pairs are reported as single-prompt positive and multi-agent negative, followed by single-prompt negative and multi-agent positive. Broken-caller counts are reported as median [interquartile range]. The Wilcoxon test uses a continuity- and tie-corrected normal approximation after removing zero paired differences. High contract/dependency risk is a pipeline-defined screening outcome, not a confirmed defect or failure probability. MA--SP denotes multi-agent minus single-prompt; pp denotes percentage points.
\end{table*}

Table~\ref{tab:rq3-generation-condition-comparison} provides no evidence of a statistically significant difference between the two generation conditions for any of the principal repository-level risk outcomes. High contract/dependency risk occurs in 660 candidates under each condition. Although four pairs are discordant, the discordance is balanced: two are classified as High risk only under single-prompt generation and two only under multi-agent generation. The exact McNemar test therefore provides no evidence of systematic asymmetry ($p=1.000$).

Detected broken-caller occurrence is descriptively lower under multi-agent generation, decreasing from 57 candidates (8.1\%) to 49 candidates (7.0\%). However, 23 pairs have a detected broken caller only under single-prompt generation, while 15 have one only under multi-agent generation. This difference is not statistically significant ($p=0.256$).

The behavioural-equivalence indicator is false for 138 single-prompt candidates (19.6\%) and 152 multi-agent candidates (21.6\%). The corresponding discordant counts are 55 and 69, and the difference is not statistically significant ($p=0.243$). The paired results therefore provide no evidence that either generation condition consistently produces fewer false behavioural-equivalence indicators.

Broken-caller counts are strongly zero-inflated. The median and upper quartile are zero under both generation conditions. The mean count is 0.641 for single-prompt candidates and 0.613 for multi-agent candidates, while the respective maxima are 35 and 42. Of the 705 pairs, 658 have identical broken-caller counts. Among the remaining pairs, 22 have a larger count under multi-agent generation and 25 have a smaller count. The Wilcoxon signed-rank test, based on the 47 non-zero paired differences, provides no evidence of a condition-level difference ($p=1.000$).

Taken together, these results do not support the conclusion that multi-agent orchestration systematically reduces the principal repository-level risk indicators examined in this study. The descriptive differences are small, vary in direction across outcomes, and are not statistically significant.

The preceding comparison examines the principal risk outcomes available across all 705 complete pairs. To determine whether the generation conditions differ in more specific risk mechanisms, we next examine the subset with usable detailed risk-reason information. Such information is available in both conditions for 115 of the 705 complete pairs. We apply two-sided exact McNemar tests to 11 mechanism indicators and use Holm correction across the complete family of comparisons.

\begin{table*}[!t]
\centering
\scriptsize
\setlength{\tabcolsep}{3.0pt}
\caption{Paired comparison of detailed risk mechanisms among the 115 pairs with usable reason information in both conditions.}
\label{tab:rq3-reason-comparison}
\resizebox{0.92\textwidth}{!}{%
\begin{tabular}{lrrrrrr}
\toprule
\rowcolor{gray!15}
\textbf{Detected mechanism}
& \textbf{Single-prompt}
& \textbf{Multi-agent}
& \textbf{MA--SP}
& \textbf{Raw $p$}
& \textbf{Holm $p$}
& \textbf{Adjusted result} \\
\midrule

Caller-used API affected
& 83 (72.2\%)
& 83 (72.2\%)
& 0.0 pp
& --
& 1.000
& Not significant \\

Public/protected signature changed
& 27 (23.5\%)
& 25 (21.7\%)
& $-1.7$ pp
& 0.500
& 1.000
& Not significant \\

Public API removed
& 6 (5.2\%)
& 2 (1.7\%)
& $-3.5$ pp
& 0.289
& 1.000
& Not significant \\

Constructor signature changed
& 19 (16.5\%)
& 18 (15.7\%)
& $-0.9$ pp
& 1.000
& 1.000
& Not significant \\

Package declaration changed
& 0 (0.0\%)
& 0 (0.0\%)
& 0.0 pp
& --
& 1.000
& Not significant \\

Large dependency delta
& 85 (73.9\%)
& 77 (67.0\%)
& $-7.0$ pp
& 0.0078
& 0.0859
& Not significant \\

Internal import/dependency change
& 9 (7.8\%)
& 11 (9.6\%)
& $+1.7$ pp
& 0.500
& 1.000
& Not significant \\

Annotation-related change
& 3 (2.6\%)
& 3 (2.6\%)
& 0.0 pp
& 1.000
& 1.000
& Not significant \\

Inheritance/interface change
& 3 (2.6\%)
& 0 (0.0\%)
& $-2.6$ pp
& 0.250
& 1.000
& Not significant \\

Any contract/API change
& 40 (34.8\%)
& 34 (29.6\%)
& $-5.2$ pp
& 0.031
& 0.313
& Not significant \\

Any dependency change
& 94 (81.7\%)
& 88 (76.5\%)
& $-5.2$ pp
& 0.031
& 0.313
& Not significant \\

\bottomrule
\end{tabular}%
}

\vspace{0.5em}
\footnotesize
\textit{Note.} The analysis includes only the 115 complete pairs with usable detailed risk reasons in both conditions. Mechanisms are not mutually exclusive. Raw $p$-values are obtained using two-sided exact McNemar tests. Holm correction is applied across all 11 mechanisms. For indicators with no discordant pairs, a value of one is included in the Holm family, while the raw test is shown as not applicable. MA--SP denotes multi-agent minus single-prompt; pp denotes percentage points.
\end{table*}

Table~\ref{tab:rq3-reason-comparison} shows several descriptive reductions under multi-agent generation. Large dependency deltas decrease by 7.0 percentage points, while any contract/API change and any dependency change each decrease by 5.2 percentage points. However, none of the 11 mechanism-level differences remains statistically significant after Holm correction.

The large dependency-delta indicator produces the smallest unadjusted $p$-value ($p=0.0078$), but its Holm-adjusted value is $p=0.0859$. The unadjusted differences in any contract/API change and any dependency change are also below 0.05, but neither remains significant after correction ($p_{\mathrm{adj}}=0.313$ for both). The analysis therefore provides no statistically supported evidence that either generation condition changes an individual risk mechanism more frequently.

The 115 analyzed pairs represent 16.3\% of the 705 complete pairs. Because detailed risk-reason information is unavailable in both conditions for most pairs, these mechanism-level findings should not be generalized to the complete paired cohort.

The condition-level analyses show no statistically significant differences between single-prompt and multi-agent generation. We next examine whether the detected risk indicators vary across the five LLM configurations. Descriptive comparisons use the complete eligible pairs available for each model. Formal repeated-measures comparisons are conducted separately within each generation condition using files with eligible candidates from all five models. This produces 108 complete five-model blocks for single-prompt generation and 149 for multi-agent generation.

\begin{table*}[!t]
\centering
\scriptsize
\setlength{\tabcolsep}{4.0pt}
\renewcommand{\arraystretch}{0.95}
\caption{Repository-level risk indicators by model and generation condition.}
\label{tab:rq3-model-descriptive}

\begin{tabular}{lrrrrrrr}
\toprule
\rowcolor{gray!15}
\multirow{2}{*}{\textbf{LLM configuration}}
& \multirow{2}{*}{\textbf{Pair $N$}}
& \multicolumn{2}{c}{\textbf{High risk}}
& \multicolumn{2}{c}{\textbf{Broken caller}}
& \multicolumn{2}{c}{
    \shortstack{\textbf{Behavioural-equivalence}\\
                \textbf{indicator false}}
} \\
\cmidrule(lr){3-4}
\cmidrule(lr){5-6}
\cmidrule(lr){7-8}
&
& \textbf{SP} & \textbf{MA}
& \textbf{SP} & \textbf{MA}
& \textbf{SP} & \textbf{MA} \\
\midrule

Claude Opus 4.8
& 150
& 94.0\% & 94.0\%
& 8.7\% & 8.0\%
& 26.0\% & 30.0\% \\

DeepSeek v4 Pro
& 148
& 93.9\% & 93.9\%
& 6.8\% & 8.1\%
& 18.9\% & 21.6\% \\

Gemini 3.1 Pro Preview
& 149
& 94.0\% & 94.6\%
& 12.1\% & 9.4\%
& 25.5\% & 23.5\% \\

GPT-5.5
& 149
& 94.0\% & 94.6\%
& 8.1\% & 7.4\%
& 15.4\% & 26.8\% \\

Qwen 3.6 Plus
& 109
& 91.7\% & 89.9\%
& 3.7\% & 0.0\%
& 9.2\% & 0.0\% \\

\midrule

Overall
& 705
& 93.6\% & 93.6\%
& 8.1\% & 7.0\%
& 19.6\% & 21.6\% \\

\bottomrule
\end{tabular}

\vspace{0.4em}
\begin{minipage}{0.88\textwidth}
\scriptsize
\textit{Note.} Rates use the complete pairs available for each model. SP denotes single-prompt generation; MA denotes multi-agent generation. Pair counts differ because some generation attempts did not produce eligible candidates. Formal model comparisons use complete five-model blocks rather than the unequal descriptive denominators shown here.
\end{minipage}
\end{table*}
Table~\ref{tab:rq3-model-descriptive} shows little separation among the model configurations in High contract/dependency risk. The observed rates range from 91.7\% to 94.0\% under single-prompt generation and from 89.9\% to 94.6\% under multi-agent generation.

Greater descriptive variation appears in detected broken-caller occurrence and false behavioural-equivalence indicators. Under single-prompt generation, broken-caller rates range from 3.7\% for Qwen to 12.1\% for Gemini. Under multi-agent generation, Qwen has no detected broken-caller cases, while the rates for the other configurations range from 7.4\% to 9.4\%. The proportion of false behavioural-equivalence indicators also varies across models. Qwen has the lowest observed proportion under both conditions and no false outcomes under multi-agent generation.

These descriptive values should not be interpreted independently as evidence that Qwen is intrinsically safer. Qwen has only 109 complete pairs because a larger number of its single-prompt attempts did not produce eligible candidates. Its observed rates are therefore conditional on successful generation and candidate eligibility.

\begin{table*}[!t]
\centering
\scriptsize
\setlength{\tabcolsep}{3.0pt}
\caption{Model-level omnibus tests within each generation condition.}
\label{tab:rq3-model-tests}
\resizebox{0.88\textwidth}{!}{%
\begin{tabular}{llrrrl}
\toprule
\rowcolor{gray!15}
\textbf{Condition}
& \textbf{Outcome}
& \textbf{Complete blocks}
& \textbf{Statistic}
& \textbf{$p$}
& \textbf{Post-hoc status} \\
\midrule

\multirow{4}{*}{Single-prompt}
& High contract/dependency risk
& 108
& Not defined
& --
& No within-block variation \\

& Broken-caller occurrence
& 108
& $Q=18.70$
& $<0.001$
& Exact McNemar with Holm \\

& Behavioural-equivalence indicator false
& 108
& $Q=29.20$
& $<0.001$
& Exact McNemar with Holm \\

& Broken-caller count
& 108
& $\chi^2_F=21.16$
& $<0.001$
& Pairwise tests not executed \\

\midrule

\multirow{4}{*}{Multi-agent}
& High contract/dependency risk
& 149
& $Q=8.57$
& 0.073
& Not applied \\

& Broken-caller occurrence
& 149
& $Q=21.89$
& $<0.001$
& Exact McNemar with Holm \\

& Behavioural-equivalence indicator false
& 149
& $Q=79.79$
& $<0.001$
& Exact McNemar with Holm \\

& Broken-caller count
& 149
& $\chi^2_F=21.92$
& $<0.001$
& Pairwise tests not executed \\

\bottomrule
\end{tabular}%
}

\vspace{0.5em}
\footnotesize
\textit{Note.} Cochran's $Q$ is used for binary outcomes, and the tie-corrected Friedman test is used for broken-caller counts. Both use asymptotic $\chi^2$ distributions with four degrees of freedom where applicable. Post-hoc exact McNemar tests are performed only after a significant Cochran test and are Holm-adjusted within each condition and outcome family. Pairwise Wilcoxon post-hoc tests were not conducted for the significant Friedman results. Consequently, no model-pair conclusion is drawn for broken-caller counts.
\end{table*}

As shown in Table~\ref{tab:rq3-model-tests}, High contract/dependency risk does not significantly distinguish the five model configurations. Under single-prompt generation, the outcome has no within-block variation: each of the five configurations is classified as High risk for the same 99 of the 108 complete blocks. Cochran's $Q$ is therefore not defined. Under multi-agent generation, the corresponding model-level difference is not statistically significant ($Q=8.57$, $p=0.073$).

Model differences are more evident for detected broken-caller occurrence. Under single-prompt generation, the omnibus test is significant ($Q=18.70$, $p<0.001$). Holm-adjusted post-hoc comparisons identify a significant difference between Gemini and Qwen ($p_{\mathrm{adj}}=0.0098$), with fewer detected broken-caller cases for Qwen. No other single-prompt model pair remains significant after adjustment.

Under multi-agent generation, broken-caller occurrence also differs across models ($Q=21.89$, $p<0.001$). Qwen has fewer detected broken-caller cases than Claude ($p_{\mathrm{adj}}=0.0044$), DeepSeek ($p_{\mathrm{adj}}=0.0044$), Gemini ($p_{\mathrm{adj}}=0.0012$), and GPT-5.5 ($p_{\mathrm{adj}}=0.0068$). The remaining model pairs do not differ significantly after adjustment.

False behavioural-equivalence indicators also differ across model configurations. Under single-prompt generation, the omnibus result is significant ($Q=29.20$, $p<0.001$). Qwen has fewer false indicator outcomes than Claude ($p_{\mathrm{adj}}=0.0029$), DeepSeek ($p_{\mathrm{adj}}=0.0105$), and Gemini ($p_{\mathrm{adj}}<0.0001$). The Qwen and GPT-5.5 comparison, as well as all comparisons not involving Qwen, is not significant after adjustment.

Under multi-agent generation, the model-level difference in false behavioural-equivalence indicators is also significant ($Q=79.79$, $p<0.001$). Qwen differs significantly from each of the other four model configurations (all $p_{\mathrm{adj}}<0.0001$). No adjusted difference is statistically significant among Claude, DeepSeek, Gemini, and GPT-5.5.

The Friedman tests additionally identify model-level heterogeneity in broken-caller counts under both single-prompt generation ($\chi^2_F=21.16$, $p<0.001$) and multi-agent generation ($\chi^2_F=21.92$, $p<0.001$). Because pairwise Wilcoxon comparisons were not conducted, these omnibus results establish only that at least one model differs within each condition. They do not identify which model pairs account for the differences.

Overall, model configuration differentiates particular detected risk channels more clearly than the choice between single-prompt and multi-agent generation. However, the observed model differences are concentrated around Qwen and should be interpreted together with its smaller number of complete pairs and its generation-failure pattern. The findings describe differences in pipeline-detected indicators and do not establish corresponding differences in semantic correctness or runtime safety.

\begin{rqanswer}
\noindent\faIcon{lightbulb}~\textbf{Answer to RQ3:}
Across the 705 complete pairs, we find no statistically significant difference between single-prompt and multi-agent generation in High contract/dependency risk, detected broken-caller occurrence, the behavioural-equivalence indicator, or broken-caller count. The comparison of 11 detailed risk mechanisms across the 115 reason-covered pairs likewise identifies no significant differences after Holm correction. Model configuration is associated with differences in detected broken-caller occurrence and false behavioural-equivalence indicators within both generation conditions, while High contract/dependency risk remains largely unchanged across models. Model configuration therefore differentiates particular detected risk channels more clearly than the choice between single-prompt and multi-agent generation. These results concern pipeline-detected indicators and do not establish differences in semantic correctness.
\end{rqanswer}

\subsection{Adoption Readiness (RQ4)}
\label{subsec:rq4-results}

RQ4 examines to what extent LLM-generated refactoring candidates appear adoption-ready under the available verification evidence. The unit of analysis is an eligible generated candidate, identified by project, file, model configuration, and generation condition. Of the 1,500 generation attempts, 1,455 produced complete candidates eligible for candidate-level analysis. These comprise 706 candidates generated under the single-prompt condition and 749 generated under the multi-agent condition.

For reporting adoption readiness, the verification evidence is summarized using four categories: \textsc{Low-risk}, \textsc{Review-needed}, \textsc{High Contract/Dependency Risk}, and \textsc{Inconclusive}. The pipeline determines overall risk by combining contract risk and observed caller-impact risk and retaining the higher risk level. All 1,365 eligible candidates assigned High overall risk also have High contract risk. We therefore report this category as \textsc{High Contract/Dependency Risk}.

\textsc{Low-risk} indicates that a candidate satisfies the pipeline's configured low-risk conditions under the available evidence. It does not establish complete behavioural equivalence. \textsc{Review-needed} indicates that a candidate does not trigger a High-risk classification but lacks sufficient evidence for the Low-risk category. \textsc{High Contract/Dependency Risk} indicates that the pipeline detects substantial contract or dependency-related risk. \textsc{Inconclusive} indicates that the available verification evidence does not support assignment to one of the other reporting categories.

Compilation evidence is unavailable for the eligible candidates. Targeted-test evidence covers only part of the multi-agent condition, and retained candidate artifacts are available for independent inspection in only a subset of cases. The categories should therefore be interpreted as conservative screening outcomes under the available evidence, rather than as complete integration decisions or independently validated judgments of semantic correctness.

\begin{table}[!t]
\centering
\scriptsize
\setlength{\tabcolsep}{2.2pt}
\caption{Adoption-readiness distribution under the available verification evidence.}
\label{tab:rq4-adoption-readiness}
\resizebox{\columnwidth}{!}{
\begin{tabular}{lrrrrr}
\toprule
\rowcolor{gray!15}
\textbf{Reporting category}
& \textbf{Count}
& \textbf{Rate}
& \textbf{Wilson 95\% CI}
& \textbf{Single-prompt}
& \textbf{Multi-agent} \\
\midrule

\textsc{Low-risk}
& 8
& 0.5\%
& 0.3--1.1\%
& 0/706 (0.0\%)
& 8/749 (1.1\%) \\

\textsc{Review-needed}
& 55
& 3.8\%
& 2.9--4.9\%
& 43/706 (6.1\%)
& 12/749 (1.6\%) \\

\textsc{High Contract/Dependency Risk}
& 1,365
& 93.8\%
& 92.5--94.9\%
& 661/706 (93.6\%)
& 704/749 (94.0\%) \\

\textsc{Inconclusive}
& 27
& 1.9\%
& 1.3--2.7\%
& 2/706 (0.3\%)
& 25/749 (3.3\%) \\

\bottomrule
\end{tabular}
}

\vspace{0.3em}
\scriptsize
\textit{Note.} Overall percentages use all 1,455 eligible candidates. Condition-specific percentages use the number of eligible candidates within each generation condition. The categories summarize evidence produced by the verification pipeline and are not independently validated semantic-correctness labels.
\end{table}

Table~\ref{tab:rq4-adoption-readiness} shows that only 8 of the 1,455 eligible candidates satisfy the pipeline's \textsc{Low-risk} conditions, corresponding to 0.5\% of the eligible cohort. A further 55 candidates (3.8\%) are classified as \textsc{Review-needed}, while 27 candidates (1.9\%) have \textsc{Inconclusive} outcomes. The dominant category is \textsc{High Contract/Dependency Risk}, which contains 1,365 candidates (93.8\%).

The proportion of candidates classified as High risk is similar across the two generation conditions. The single-prompt condition contains 661 of 706 High-risk candidates (93.6\%), while the multi-agent condition contains 704 of 749 (94.0\%). All eight \textsc{Low-risk} candidates occur in the multi-agent condition. \textsc{Review-needed} candidates are more frequent under single-prompt generation, whereas \textsc{Inconclusive} outcomes are more frequent under multi-agent generation.

These descriptive differences should be interpreted cautiously because downstream verification coverage differs between the two conditions. In particular, targeted-test evidence is available only for multi-agent candidates. The adoption-readiness distributions therefore combine structural evidence available for all eligible candidates with verification evidence available only for a subset.

The concentration of candidates in the High-risk category reflects the decision rules implemented by the verification pipeline. High contract risk can be triggered by configured structural conditions, including contract-change indicators and a dependency delta meeting the implemented threshold. The 93.8\% result therefore represents the prevalence of pipeline-defined High contract/dependency risk within the eligible cohort. It does not mean that 93.8\% of the candidates were independently demonstrated to be incorrect, uncompilable, or behaviourally unsafe.

\begin{table*}[!t]
\centering
\footnotesize
\setlength{\tabcolsep}{7pt}
\renewcommand{\arraystretch}{0.95}
\caption{Coverage and outcomes of the adoption-readiness evidence.}
\label{tab:rq4-verification-coverage}

\begin{tabular}{p{0.62\textwidth}rr}
\toprule
\rowcolor{gray!15}
\textbf{Evidence item}
& \textbf{Count}
& \textbf{Rate} \\
\midrule

\multicolumn{3}{l}{\textit{Structural and behavioural evidence ($n=1{,}455$ eligible candidates)}} \\
High contract risk
& 1,365
& 93.8\% \\

Medium contract risk
& 90
& 6.2\% \\

Broken caller detected
& 106
& 7.3\% \\

Behavioural-equivalence indicator: true
& 1,165
& 80.1\% \\

Behavioural-equivalence indicator: false
& 290
& 19.9\% \\

\addlinespace[2pt]
\multicolumn{3}{l}{\textit{Targeted-test availability ($n=1{,}455$ eligible candidates)}} \\
Test eligible
& 395
& 27.1\% \\

Not test eligible or unavailable
& 1,060
& 72.9\% \\

\addlinespace[2pt]
\multicolumn{3}{l}{\textit{Targeted-test conclusiveness ($n=395$ test-eligible candidates)}} \\
Conclusive outcome
& 65
& 16.5\% \\

Inconclusive outcome
& 330
& 83.5\% \\

\addlinespace[2pt]
\multicolumn{3}{l}{\textit{Conclusive targeted-test results ($n=65$)}} \\
Passed
& 60
& 92.3\% \\

Regression detected
& 5
& 7.7\% \\

\addlinespace[2pt]
\multicolumn{3}{l}{\textit{Artifact availability ($n=1{,}455$ eligible candidates)}} \\
Candidate artifact available
& 187
& 12.9\% \\

Candidate source file unavailable
& 1,268
& 87.1\% \\

\bottomrule
\end{tabular}

\vspace{0.4em}
\begin{minipage}{0.84\textwidth}
\footnotesize
\textit{Note.} Structural and behavioural indicators are available for all eligible candidates. Targeted-test evidence is available only for 395 multi-agent candidates. Compilation evidence is unavailable. Passing the available targeted tests does not establish complete semantic equivalence.
\end{minipage}
\end{table*}
Table~\ref{tab:rq4-verification-coverage} summarizes the evidence underlying the adoption-readiness categories. Structural risk assessment is available for all 1,455 eligible candidates. Of these, 1,365 have High contract risk and 90 have Medium contract risk. At least one broken downstream caller is detected for 106 candidates, corresponding to 7.3\% of the eligible cohort.

Across all 1,500 generation attempts, 126 records contain detected broken-caller evidence. After restricting RQ4 to the 1,455 complete candidates eligible for candidate-level analysis, 106 candidates retain this evidence. The two counts therefore describe different analytical populations: 126 applies to all generation attempts, whereas 106 applies to the eligible-candidate cohort used in RQ4.

The behavioural-equivalence indicator is true for 1,165 candidates (80.1\%) and false for 290 candidates (19.9\%). This indicator provides one form of automated behavioural-risk evidence, but it should not be interpreted as a substitute for compilation, comprehensive testing, or manual semantic assessment.

Targeted-test evidence has more limited coverage. Only 395 of the 1,455 eligible candidates are test eligible, corresponding to 27.1\% of the eligible cohort, and all 395 belong to the multi-agent condition. Of these candidates, 65 produce conclusive targeted-test outcomes. Sixty of the 65 conclusive outcomes pass the targeted tests, while five exhibit test regressions. The remaining 330 test-eligible candidates have inconclusive targeted-test outcomes.

The 60 passing outcomes provide positive evidence for the corresponding candidates, but they should not be generalized to the complete cohort. They represent 92.3\% of the 65 conclusive outcomes, not 92.3\% of all candidates or all test-eligible candidates. Furthermore, passing the available targeted tests does not establish complete behavioural equivalence because the tests may not exercise every behaviour affected by a candidate.

Compilation evidence is unavailable and therefore cannot provide an additional verification layer. Candidate artifacts are available for independent inspection in 187 cases, corresponding to 12.9\% of the eligible cohort, and unavailable in the remaining 1,268 cases. These coverage limitations constrain candidate-level adoption claims and prevent the study from estimating the precision of the High-risk classification.

\begin{table*}[!t]
\centering
\small
\setlength{\tabcolsep}{2.5pt}
\caption{Adoption-readiness distribution by LLM configuration.}
\label{tab:rq4-model-adoption}
\resizebox{0.90\textwidth}{!}{
\begin{tabular}{lrrrrr}
\toprule
\rowcolor{gray!15}
\textbf{LLM configuration}
& \textbf{Eligible $n$}
& \textbf{\textsc{Low-risk}}
& \textbf{\textsc{Review-needed}}
& \textbf{\textsc{High risk}}
& \textbf{\textsc{Inconclusive}} \\
\midrule

Claude Opus 4.8
& 300
& 1 (0.3\%)
& 9 (3.0\%)
& 282 (94.0\%)
& 8 (2.7\%) \\

DeepSeek v4 Pro
& 298
& 2 (0.7\%)
& 11 (3.7\%)
& 280 (94.0\%)
& 5 (1.7\%) \\

Gemini 3.1 Pro Preview
& 299
& 2 (0.7\%)
& 11 (3.7\%)
& 282 (94.3\%)
& 4 (1.3\%) \\

GPT-5.5
& 299
& 1 (0.3\%)
& 11 (3.7\%)
& 282 (94.3\%)
& 5 (1.7\%) \\

Qwen 3.6 Plus
& 259
& 2 (0.8\%)
& 13 (5.0\%)
& 239 (92.3\%)
& 5 (1.9\%) \\

\bottomrule
\end{tabular}
}

\vspace{0.5em}
\footnotesize
\textit{Note.} Model denominators differ because some generation attempts did not produce complete candidates eligible for candidate-level analysis. The results are descriptive and should not be interpreted as a ranking of model correctness.
\end{table*}

Table~\ref{tab:rq4-model-adoption} shows that \textsc{High Contract/Dependency Risk} dominates the adoption-readiness distribution for every LLM configuration. The proportion ranges from 92.3\% for Qwen 3.6 Plus to 94.3\% for Gemini 3.1 Pro Preview and GPT-5.5. Claude Opus 4.8 and DeepSeek v4 Pro each have a High-risk proportion of 94.0\%.

Only one or two candidates per model satisfy the \textsc{Low-risk} conditions. Because the complete eligible cohort contains only eight Low-risk candidates, these counts do not provide sufficient evidence for substantive model-ranking claims. The small differences in High-risk proportions should also be interpreted in light of the structural-risk rules and the unequal numbers of eligible candidates across model configurations.

Overall, very few candidates satisfy the pipeline's \textsc{Low-risk} conditions, while most trigger the \textsc{High Contract/Dependency Risk} classification. However, the available evidence does not demonstrate that most candidates are semantically incorrect. Instead, the classifications provide a conservative basis for screening generated candidates and prioritizing additional repository-level verification before adoption.

\begin{rqanswer}
\noindent\faIcon{lightbulb}~\textbf{Answer to RQ4:}
Under the available verification evidence, only 8 of 1,455 eligible candidates (0.5\%) satisfy the pipeline's \textsc{Low-risk} conditions. Most candidates, comprising 1,365 cases (93.8\%), are classified as having \textsc{High Contract/Dependency Risk}. A further 55 candidates (3.8\%) require additional review, and 27 candidates (1.9\%) have inconclusive adoption-readiness outcomes. Among the 65 candidates with conclusive targeted-test outcomes, 60 pass and five exhibit regressions. However, targeted tests cover only a restricted multi-agent subset, compilation evidence is unavailable, and retained candidate artifacts are available for inspection in only a subset of cases. The results therefore show that most generated refactorings require further repository-level verification before adoption. They do not establish that 93.8\% of the candidates are semantically incorrect.
\end{rqanswer}

\subsection{Factors Associated with Screening and Observed-Impact Outcomes (RQ5)}
\label{subsec:rq5-results}

RQ5 examines which source-file characteristics and generated-change indicators are associated with the pipeline's contract/dependency-risk and observed-impact outcomes. The analysis covers 1,455 eligible candidates generated from 150 source files, comprising 706 single-prompt candidates and 749 multi-agent candidates.

We analyze three outcomes separately. The first is the pipeline-defined High contract/dependency-risk classification, which occurs for 1,365 of 1,455 candidates (93.8\%). The second is detected broken-caller evidence, which occurs for 106 candidates (7.3\%). The third is a false behavioural-equivalence indicator, which occurs for 290 candidates (19.9\%). High contract/dependency risk is a rule-based screening classification, broken-caller evidence is a static observed-impact indicator, and the behavioural-equivalence indicator is a heuristic measure. None of these outcomes independently establishes semantic failure.

The primary analysis uses the source file as its unit. Each of the 150 source files contributes one observation, and each outcome is represented by the proportion of eligible candidates generated for that file that exhibit the corresponding outcome. This approach reduces the dependence created by generating multiple candidates from the same source file. It does not completely account for project-level clustering, so the results are interpreted as exploratory rather than causal or predictive.

For continuous file characteristics, we use Spearman's rank correlation between the characteristic and the file-level outcome proportion. For binary characteristics, we use Cliff's $\delta$ to compare files with and without the characteristic. Holm correction is applied separately within each outcome family. Candidate-level analyses are used only as sensitivity checks because treating repeated candidates from the same file as independent observations can overstate statistical precision.

\begin{table*}[!t]
\centering
\scriptsize
\setlength{\tabcolsep}{3.0pt}
\caption{File-level associations between pre-generation source-file characteristics and the three RQ5 outcomes ($N=150$ files).}
\label{tab:rq5-file-associations}
\resizebox{0.94\textwidth}{!}{
\begin{tabular}{llll}
\toprule
\rowcolor{gray!15}
\textbf{Source-file characteristic}
& \textbf{High-risk share}
& \textbf{Broken-caller share}
& \textbf{False behavioural-equivalence share} \\
\midrule

LOC
& $\rho=0.317$, $p_{\mathrm{Holm}}<0.001$
& $\rho=0.263$, $p_{\mathrm{Holm}}=0.006$
& $\rho=0.369$, $p_{\mathrm{Holm}}<0.001$ \\

Baseline smell count
& $\rho=0.154$, $p_{\mathrm{Holm}}=0.119$
& $\rho=0.261$, $p_{\mathrm{Holm}}=0.006$
& $\rho=0.446$, $p_{\mathrm{Holm}}<0.001$ \\

Fan-in
& $\rho=0.341$, $p_{\mathrm{Holm}}<0.001$
& $\rho=0.324$, $p_{\mathrm{Holm}}<0.001$
& $\rho=0.112$, $p_{\mathrm{Holm}}=0.521$ \\

Fan-out
& $\rho=0.361$, $p_{\mathrm{Holm}}<0.001$
& $\rho=0.181$, $p_{\mathrm{Holm}}=0.080$
& $\rho=0.224$, $p_{\mathrm{Holm}}=0.029$ \\

Public/protected API count
& $\rho=0.305$, $p_{\mathrm{Holm}}<0.001$
& $\rho=0.282$, $p_{\mathrm{Holm}}=0.003$
& $\rho=0.310$, $p_{\mathrm{Holm}}<0.001$ \\

Caller-used API count
& $\rho=0.341$, $p_{\mathrm{Holm}}<0.001$
& $\rho=0.404$, $p_{\mathrm{Holm}}<0.001$
& $\rho=0.232$, $p_{\mathrm{Holm}}=0.026$ \\

Downstream-caller count
& $\rho=0.341$, $p_{\mathrm{Holm}}<0.001$
& $\rho=0.324$, $p_{\mathrm{Holm}}<0.001$
& $\rho=0.111$, $p_{\mathrm{Holm}}=0.521$ \\

Baseline dependency-risk score
& $\rho=0.374$, $p_{\mathrm{Holm}}<0.001$
& $\rho=0.351$, $p_{\mathrm{Holm}}<0.001$
& $\rho=0.333$, $p_{\mathrm{Holm}}<0.001$ \\

Inheritance present
& $\delta=0.024$, $p_{\mathrm{Holm}}=0.574$
& $\delta=0.106$, $p_{\mathrm{Holm}}=0.176$
& $\delta=0.079$, $p_{\mathrm{Holm}}=0.521$ \\

Interface present
& $\delta=0.102$, $p_{\mathrm{Holm}}=0.056$
& $\delta=-0.005$, $p_{\mathrm{Holm}}=0.937$
& $\delta=0.185$, $p_{\mathrm{Holm}}=0.157$ \\

\bottomrule
\end{tabular}
}

\vspace{0.4em}
\scriptsize
\textit{Note.} Continuous characteristics are assessed using Spearman's $\rho$. Binary inheritance and interface characteristics are assessed using Cliff's $\delta$. Reported $p$-values are Holm adjusted within each outcome family. Fan-in, caller-used API count, and downstream-caller count are closely related exposure measures and should not be interpreted as independent effects.
\end{table*}

Table~\ref{tab:rq5-file-associations} shows that seven of the ten source-file characteristics have Holm-supported associations with the proportion of High-risk candidates. The strongest association is observed for baseline dependency-risk score ($\rho=0.374$, $p_{\mathrm{Holm}}<0.001$), followed by fan-out ($\rho=0.361$, $p_{\mathrm{Holm}}<0.001$). Fan-in, caller-used API count, downstream-caller count, LOC, and public/protected API count also have positive associations with the High-risk share. Baseline smell count, inheritance presence, and interface presence are not supported after Holm correction.

These High-risk associations must be interpreted in relation to the outcome definition. High contract/dependency risk is assigned using structural rules that include contract and dependency changes. The positive associations therefore show that files with greater pre-generation exposure receive larger proportions of pipeline-defined High-risk classifications. They do not provide independent evidence that these files produce more semantically incorrect refactorings.

Seven of the ten file characteristics also have Holm-supported associations with broken-caller share. Caller-used API count has the strongest association ($\rho=0.404$, $p_{\mathrm{Holm}}<0.001$). Baseline dependency-risk score, fan-in, downstream-caller count, public/protected API count, LOC, and baseline smell count are also positively associated with broken-caller share. Fan-out, inheritance presence, and interface presence are not supported after Holm correction.

The broken-caller results provide more direct observed-impact evidence than the High-risk classification, although they remain based on static caller analysis rather than compilation or execution. The findings indicate that files with more caller-used APIs and greater inbound dependency exposure tend to produce a larger proportion of candidates with detected caller impact.

Six of the ten source-file characteristics have Holm-supported associations with the proportion of candidates for which the behavioural-equivalence indicator is false. Baseline smell count has the strongest association ($\rho=0.446$, $p_{\mathrm{Holm}}<0.001$). LOC, baseline dependency-risk score, public/protected API count, caller-used API count, and fan-out are also positively associated with this outcome. Fan-in, downstream-caller count, inheritance presence, and interface presence are not supported after Holm correction.

All Holm-supported file-level associations are positive and have small to medium effect sizes. Inheritance and interface presence do not have Holm-supported file-level associations with any of the three outcomes. The strongest associations differ across outcomes, further supporting their separate analysis rather than treating them as interchangeable measures of candidate correctness.

The candidate-level sensitivity analysis produces a broader set of statistically supported factors. For High contract/dependency risk, all ten characteristics are supported at the candidate level, compared with seven at the file level. For broken-caller evidence, nine are supported at the candidate level, compared with seven at the file level. For the false behavioural-equivalence indicator, all ten are supported at the candidate level, compared with six at the file level. The additional candidate-level associations involve baseline smell count, inheritance presence, interface presence, fan-out, fan-in, and downstream-caller count, depending on the outcome. Because these tests treat repeated candidates from the same source file as independent, the additional findings may reflect clustering-related precision inflation. We therefore base the principal interpretation on the file-level results.

The High-risk classification can also be decomposed according to the risk-reason flags recorded by the pipeline. Machine-readable risk reasons are available for 864 of the 1,455 eligible candidates (59.4\%). Coverage includes all 749 multi-agent candidates but only 115 of 706 single-prompt candidates. Among these 864 candidates, 810 receive a High-risk classification under the implemented dependency-delta threshold of five. Table~\ref{tab:rq5-risk-mechanisms} reports the mechanisms recorded for these 810 High-risk candidates.

\begin{table*}[!t]
\centering
\scriptsize
\setlength{\tabcolsep}{3.5pt}
\caption{Risk-rule mechanisms among High-risk candidates with machine-readable risk reasons.}
\label{tab:rq5-risk-mechanisms}
\resizebox{0.88\textwidth}{!}{
\begin{tabular}{lrrr}
\toprule
\rowcolor{gray!15}
\textbf{Risk-rule mechanism}
& \textbf{High-risk candidates}
& \textbf{Share of 810 High-risk candidates}
& \textbf{Sole High-producing mechanism} \\
\midrule

Caller-used API affected
& 597
& 73.7\%
& 114 \\

Public/protected signature changed
& 231
& 28.5\%
& 0 \\

Public API removed
& 82
& 10.1\%
& 0 \\

Constructor signature changed
& 138
& 17.0\%
& 6 \\

Package changed
& 0
& 0.0\%
& 0 \\

Large dependency delta ($\mathit{depDelta}\geq 5$)
& 662
& 81.7\%
& 150 \\

Annotation changed
& 78
& 9.6\%
& 2 \\

Inheritance or interface changed
& 28
& 3.5\%
& 0 \\

\bottomrule
\end{tabular}
}

\vspace{0.4em}
\scriptsize
\textit{Note.} Risk-reason flags overlap because a candidate can trigger more than one rule. The denominator is the 810 High-risk candidates among the 864 candidates with usable machine-readable risk reasons. These mechanisms describe how the pipeline assigned High risk. They are not independent predictors or causes of candidate failure.
\end{table*}

Large dependency delta is the most frequent recorded mechanism, occurring for 662 of the 810 High-risk candidates with usable risk reasons (81.7\%). It is also the sole High-producing mechanism for 150 candidates. Caller-used API impact occurs for 597 High-risk candidates (73.7\%) and is the sole High-producing mechanism for 114 candidates. Constructor signature change is the sole High-producing mechanism for six candidates, and annotation change is the sole mechanism for two candidates. The remaining mechanisms do not independently produce a High classification in this subset.

These results explain why High-risk classifications are common, but they do not validate the precision of the classification. The mechanisms are components of the implemented decision rule and therefore cannot be treated as external predictors of the same outcome. The threshold-sensitivity analysis also shows that the dependency-delta rule materially affects the number of High-risk candidates. Within the 864 candidates with usable risk reasons, the High-risk count decreases from 810 at a threshold of five to 714 at a threshold of 20.

Post-generation change metrics have uneven coverage. Churn, lines added, and lines removed are available for 778 of 1,455 candidates (53.5\%). These data cover all 706 single-prompt candidates but only 72 of 749 multi-agent candidates. Consequently, the available-case results for these metrics are dominated by the single-prompt condition and cannot be generalized to the complete eligible cohort. None of the three metrics has a Holm-supported association with any of the three RQ5 outcomes. Missing telemetry values are retained as missing rather than converted to zero.

Smell delta is available for all 1,455 eligible candidates. It does not have a Holm-supported association with the High-risk classification. It is associated with broken-caller evidence, with Cliff's $\delta=-0.54$, and with a false behavioural-equivalence indicator, with Cliff's $\delta=-0.59$. The negative effect sizes indicate that candidates exhibiting these outcomes tend to have lower smell-delta values under the study's encoding. These findings show that local smell changes and repository-level impact indicators capture different properties of a generated refactoring. They should not be interpreted as evidence that smell reduction causes caller impact or behavioural change.

The RQ5 results should be interpreted in light of several limitations. The precision of the High-risk classification is unknown, and the classification is defined by configured structural rules, including the dependency-delta threshold. Risk-reason coverage is incomplete and uneven across the two generation conditions, compilation evidence is unavailable, and the regular-expression-based extraction of risk reasons may introduce measurement error. File-level aggregation reduces dependence among repeated candidate observations but does not fully model project-level dependence. The reported associations are therefore exploratory and should not be interpreted as causal, predictive, or as estimates of real-world failure probability.

\begin{rqanswer}
\noindent\faIcon{lightbulb}~\textbf{Answer to RQ5:}
Source files with greater repository exposure tend to have higher shares of pipeline-defined High contract/dependency risk and detected broken-caller evidence. At the file level, baseline dependency-risk score has the strongest association with High-risk share ($\rho=0.374$, $p_{\mathrm{Holm}}<0.001$), caller-used API count has the strongest association with broken-caller share ($\rho=0.404$, $p_{\mathrm{Holm}}<0.001$), and baseline smell count has the strongest association with a false behavioural-equivalence indicator ($\rho=0.446$, $p_{\mathrm{Holm}}<0.001$). Seven of the ten source-file characteristics are associated with High-risk share, seven are associated with broken-caller share, and six are associated with false behavioural-equivalence share after Holm correction. These findings identify exploratory associations with the pipeline's screening and observed-impact indicators. They do not establish causal relationships or estimate the probability that a generated candidate is semantically incorrect.
\end{rqanswer}

\subsection{Observed Risk Mechanisms}
\label{subsec:observed-risk-mechanisms}

The preceding RQs show that pipeline-defined High contract/dependency risk is widespread among the eligible candidates. To avoid treating the aggregate classification as a single failure mode, this subsection summarizes the mechanisms recorded by the verification pipeline. These mechanisms describe how candidates trigger the implemented screening rules or exhibit detected caller impact. They should not be interpreted as independent causes of semantic failure.

Machine-readable risk-reason information is available for 864 of the 1,455 eligible candidates (59.4\%). This subset includes all 749 multi-agent candidates but only 115 of 706 single-prompt candidates. The mechanism counts are therefore coverage-limited descriptive results and should not be generalized to the complete single-prompt cohort.

The first mechanism is \emph{dependency-structure change}. Among the 864 candidates with usable risk-reason information, 716 candidates (82.9\%) exhibit at least one recorded dependency-change indicator. A large dependency delta, defined by the implemented threshold of five or more, occurs in 662 candidates (76.6\%), while 54 candidates (6.3\%) exhibit an import-change indicator. Large dependency delta is the most frequent individual mechanism in the reason-covered subset. It indicates that many generated candidates alter the dependency structure detected for the edited file. However, its prevalence is sensitive to the configured threshold. In the threshold analysis, the number of High-risk candidates in the reason-covered subset decreases from 810 at a threshold of five to 714 at a threshold of 20.

The second mechanism is \emph{caller-used API change}. Among the same 864 candidates, 597 candidates (69.1\%) change or affect an API identified as being used by downstream callers. This mechanism connects a local edit with repository-level exposure because changes to a caller-used API may require coordinated updates or additional verification in dependent files. However, the indicator is produced by the study's extraction and comparison pipeline. It should therefore be interpreted as a detected caller-facing change rather than as confirmed caller failure.

The third mechanism is \emph{contract and public-surface change}. At least one contract-related risk-reason flag is present for 333 of the 864 reason-covered candidates (38.5\%). Public or protected signature changes occur in 231 candidates (26.7\%), constructor signature changes occur in 138 candidates (16.0\%), and public API removal occurs in 82 candidates (9.5\%). These categories overlap because one candidate can trigger multiple mechanisms. No package-change cases are recorded in the reason-covered subset.

These contract-related mechanisms are relevant because public and protected members can form repository-facing interfaces. Even when a change is intentional, its adoption may require verification of affected callers, subclasses, implementations, or dependent components. The mechanism counts do not show that every detected contract change is incorrect. They show that a generated candidate differs from its source file in a way that the pipeline identifies as contract relevant.

The fourth mechanism is \emph{framework and structural change}. Annotation-related changes occur in 78 of the 864 reason-covered candidates (9.0\%), while inheritance or implemented-interface changes occur in 28 candidates (3.2\%). These mechanisms are less frequent than dependency and caller-used API changes. Nevertheless, they can be important because annotations, inheritance relationships, and interface implementations may encode framework assumptions, lifecycle behaviour, dependency injection, serialization rules, or polymorphic contracts. Such changes may affect repository behaviour without producing an immediately detected broken caller.

The fifth mechanism is \emph{detected downstream caller impact}. Across the complete eligible cohort, 106 of 1,455 candidates (7.3\%) have at least one detected broken downstream caller. The remaining 1,349 candidates (92.7\%) have no detected broken caller. The number of affected callers is concentrated in a small number of candidates. The largest observed candidate-level count is 42 broken callers, followed by cases with 40 and 35 broken callers.

Direct broken-caller evidence is therefore considerably less frequent than pipeline-defined High contract/dependency risk. This difference shows that the two measures capture different constructs. A High-risk classification can be triggered by structural rules even when no broken caller is detected. Conversely, a nonzero broken-caller count provides more direct static evidence that a generated change may affect downstream code, although it does not establish compilation or runtime failure.

Taken together, the pipeline-defined risk classifications arise through several overlapping channels: dependency changes, caller-used API changes, public contract changes, framework-related changes, inheritance or interface changes, and detected caller impact. Large dependency delta and caller-used API impact are the most frequent recorded mechanisms in the reason-covered subset.

These findings reinforce the distinction between local improvement and repository-level adoption readiness. A generated candidate may reduce PMD-detected smells or simplify the edited file while also changing repository-facing structures. Such a candidate may require additional caller analysis, compilation, targeted testing, or human review before adoption. At the same time, the mechanism counts should not be interpreted as estimates of semantic-failure prevalence because risk-reason coverage is incomplete, the High-risk rules are configuration dependent, and the precision of the individual indicators has not been established through systematic manual validation.

%% file: sections/05-discussion.tex
\section{Discussion and Implications}
\label{sec:discussion}

The results show that LLM-based refactoring should be studied not only as a local code-improvement problem, but also as a repository-level verification problem. The selected files commonly expose APIs, dependencies, and caller relationships through which a local change can affect other parts of the repository. The generated candidates also trigger multiple structural-screening and observed-impact indicators. However, these indicators should not be combined into a single claim that most candidates are semantically incorrect.

The study distinguishes three principal forms of evidence. Pipeline-defined High contract/dependency risk is a rule-based structural-screening outcome. Broken-caller evidence identifies detected static impact on downstream callers. The behavioural-equivalence indicator provides a separate heuristic signal concerning behaviour preservation. These measures have different prevalence, coverage, and interpretations. Compilation evidence is unavailable, targeted-test evidence covers only a subset of multi-agent candidates, and systematic manual validation has not established the precision of the High-risk classification. The central contribution is therefore a repository-level account of where generated refactorings require further verification, rather than an estimate of how frequently LLM-generated refactorings fail.

\subsection{Repository-Level Safety}
\label{subsec:discussion-repository-safety}

Refactoring is traditionally defined as restructuring software without changing its observable behaviour~\cite{fowler2018refactoring}. Prior refactoring research has therefore emphasized behaviour preservation, tool support, and the effects of refactoring on the surrounding software process~\cite{mens2004survey}. This preservation requirement becomes especially important for LLM-generated refactorings because a generated edit may appear locally plausible while modifying contracts, dependencies, or structures used elsewhere in the repository.

Recent literature shows that LLMs are increasingly applied to code generation, maintenance, repair, and quality improvement~\cite{hou2024large}. Repository-level coding research further demonstrates that realistic software changes often require context and planning beyond an isolated code fragment~\cite{bairi2024codeplan}. Our findings extend this repository-level perspective to refactoring verification. Repository context is relevant not only when generating a candidate, but also when determining what evidence is required before that candidate can be considered for adoption.

The files selected in this study are not isolated units. Their exposure profiles include outbound dependencies, public or protected APIs, inbound references, caller-used APIs, downstream callers, inheritance relationships, and framework-sensitive constructs. Exposure is also uneven. Many files have relatively limited inbound exposure, while a smaller number of highly connected files have substantially larger repository-facing surfaces. Consequently, apparently similar file-level refactoring tasks can present very different opportunities for ripple effects.

The RQ5 analysis supports this interpretation. At the file level, baseline dependency-risk score has the strongest association with the proportion of candidates receiving a High contract/dependency-risk classification. Caller-used API count has the strongest association with detected broken-caller share, while baseline smell count has the strongest association with the proportion of false behavioural-equivalence indicators. These outcomes should not be treated as interchangeable, but they consistently show that characteristics of the source file influence the amount and type of verification evidence produced after generation.

The broader implication is that the relevant unit of concern is the edited file within its repository context. Local quality indicators remain useful for evaluating the intended benefit of a refactoring, but they do not establish that a candidate is ready for repository integration. Repository-level verification should therefore complement measures such as smell reduction, complexity, and maintainability proxies. This is especially important for files with caller-used APIs, large dependency surfaces, or broad public and protected interfaces.

\subsection{Local Improvement and Adoption Readiness}
\label{subsec:discussion-improvement-adoption}

Prior studies provide evidence that LLMs can generate useful refactoring suggestions and improve local code-quality indicators. Empirical work on ChatGPT-based refactoring has examined whether generated transformations preserve behaviour and improve quality attributes in Java code~\cite{depalma2024exploring}. Other research reports that code-specialized LLMs can reduce detected smells and that prompting strategies can influence smell reduction and test outcomes~\cite{cordeiro2026capability}. Our findings do not contradict this potential. Instead, they show that local improvement and repository-level adoption readiness are distinct empirical constructs.

This distinction is visible in the generated-change analysis. Smell delta is not associated with the pipeline-defined High-risk classification after correction for multiple comparisons. However, it is associated with broken-caller evidence and false behavioural-equivalence indicators. These associations have negative effect sizes under the study's encoding, indicating that candidates exhibiting these outcomes tend to have lower smell-delta values. Changes in local smell counts and repository-level impact indicators can therefore occur together. Local improvement does not provide sufficient evidence of repository-level preservation.

The adoption-readiness analysis demonstrates the same distinction. Only 8 of the 1,455 eligible candidates (0.5\%) satisfy the pipeline's Low-risk conditions, while 1,365 candidates (93.8\%) receive the High contract/dependency-risk classification. These values describe the output of a conservative screening rule. They do not demonstrate that 93.8\% of the candidates are defective. The dependency-delta threshold contributes directly to the High-risk classification. Within the 864 candidates with usable risk reasons, increasing the threshold from five to 20 reduces the High-risk count from 810 to 714. The observed prevalence is therefore partly a property of the configured screening policy.

The targeted-test evidence provides an important complementary result. Among 65 candidates with conclusive targeted-test outcomes, 60 pass and five exhibit regressions. Fifty of the 60 passing candidates also receive a High-risk classification. This result shows that structural screening and targeted tests capture different properties of a candidate. However, the 65 conclusive outcomes represent a restricted subset of the 395 test-eligible multi-agent candidates and a much smaller subset of the complete eligible cohort. Passing the available targeted tests also does not establish general behavioural equivalence. The result should therefore be interpreted as positive but limited evidence, not as an estimate of the precision of the High-risk classification.

These findings support a separation between improvement evidence and adoption evidence. Improvement evidence concerns whether a candidate achieves its intended local benefit. Adoption evidence concerns whether contracts, callers, dependencies, build behaviour, and relevant tests remain acceptable. A candidate may provide useful local improvement while still requiring repository-level review. Conversely, a structural-risk signal may identify an edit that needs inspection without proving that the edit is semantically incorrect.

\subsection{Role of Dependency Context}
\label{subsec:discussion-dependency-context}

Repository-level context is increasingly recognized as important for LLM-based software engineering because realistic changes depend on information distributed across files, dependencies, and previous edits. CodePlan, for example, treats repository-level coding as a planning problem in which edit steps are supported by repository context, prior changes, and task-specific instructions~\cite{bairi2024codeplan}. Repository-level code-generation research also indicates that dependency context can influence executability, functional correctness, and dependency use~\cite{hai2025impacts}. These studies support the general argument that repository context matters for LLM-based code tasks.

Our findings refine this argument for refactoring. Dependency context serves two distinct purposes. It can inform generation by showing the model which APIs, callers, and dependencies should be preserved. It can also support verification by revealing whether the resulting candidate changes those structures. Providing context during generation does not remove the need for post-generation verification.

Machine-readable risk reasons are available for 864 of the 1,455 eligible candidates. Within this coverage-limited subset, 716 candidates (82.9\%) exhibit at least one recorded dependency-change indicator, 662 candidates (76.6\%) exhibit a large dependency delta at the implemented threshold, and 597 candidates (69.1\%) affect a caller-used API. At least one contract-related flag is present for 333 candidates (38.5\%). These indicators overlap because a candidate may trigger several rules.

These values show that dependency-aware generation does not guarantee structural preservation. However, they should be interpreted carefully. Risk-reason coverage includes all 749 multi-agent candidates but only 115 of the 706 single-prompt candidates. The reason-tag results therefore describe the covered subset and cannot be generalized uniformly across both conditions. In addition, the indicators are produced by the implemented extraction and comparison pipeline. Their precision has not been established through systematic manual or parser-based validation.

The dependency threshold also illustrates why generation context and verification policy should remain conceptually separate. A candidate can receive the same prompt context while its final classification changes under a different screening threshold. The threshold does not alter the generated code. It alters how the resulting dependency change is categorized. Systems should therefore report both the observed dependency delta and the policy used to interpret it, rather than presenting the resulting category as an intrinsic property of the candidate.

The practical implication is that dependency context should be part of a risk-aware workflow, not treated as a guarantee of preservation. Context can guide the LLM, while post-generation analysis can identify caller-used API changes, dependency deltas, contract changes, annotation changes, inheritance changes, and detected caller impact. These signals should be exposed separately and combined with build and behavioural evidence when such evidence is available.

\subsection{Generation-Condition and Model Effects}
\label{subsec:discussion-generation-model}

Recent work on agentic and repository-level LLM refactoring reflects a broader movement from single-prompt generation toward structured workflows involving planning, retrieval, validation, and specialized agents. RefAgent frames automatic refactoring as a multi-agent process and evaluates generated refactorings, opportunity identification, and the contribution of individual agents~\cite{oueslati2026refagent}. RepoAI similarly presents repository-level refactoring as a multi-agent orchestration problem supported by retrieval and cross-family validation~\cite{chondamrongkul2026repoai}. These approaches suggest that orchestration can introduce structure beyond a single prompt.

Our paired analysis does not support a general claim that one generation condition produces candidates with lower repository-level risk. Among the 705 complete single-prompt and multi-agent pairs, 660 candidates in each condition receive the High contract/dependency-risk classification, corresponding to 93.6\% in both conditions. Broken-caller evidence occurs for 57 single-prompt candidates and 49 multi-agent candidates, but the paired difference is not statistically significant. The behavioural-equivalence indicator is false for 138 single-prompt candidates and 152 multi-agent candidates, and this difference is also not statistically significant.

The comparison of broken-caller counts leads to the same conclusion. Broken-caller counts are strongly zero-inflated, with a median of zero in both conditions. The paired Wilcoxon analysis does not identify evidence of a difference. The findings therefore do not support the claim that multi-agent generation reduces caller impact or that single-prompt generation produces fewer false behavioural-equivalence indicators.

The reason-tag comparison has more limited coverage. Only 115 of the 705 complete pairs have usable risk reasons for both conditions. No mechanism-level comparison remains statistically significant after Holm correction. Large dependency delta produces the smallest unadjusted $p$-value, but its Holm-adjusted value is 0.086. The available evidence therefore does not support claims that one condition systematically increases or decreases dependency changes, contract changes, or caller-used API effects.

Model-level results show greater variation for particular indicators. Within the complete model blocks, model differences are observed for broken-caller occurrence and false behavioural-equivalence indicators under both generation conditions. Several statistically significant pairwise comparisons involve Qwen 3.6 Plus. However, Qwen has fewer complete pairs because more of its corresponding generation attempts did not produce eligible candidates. Within the complete paired cohort, its multi-agent candidates have no detected broken-caller cases and no false behavioural-equivalence indicators. These values should be reported as observed differences rather than interpreted as evidence that Qwen is categorically more correct or safer than the other configurations.

The broader implication is that models and generation conditions should be compared through indicator profiles rather than a single safety ranking. A model may differ in broken-caller occurrence while showing similar High-risk prevalence. A generation workflow may change the distribution of specific signals without producing a statistically supported difference in the principal paired outcomes. Reporting these distinctions is more informative than assigning a model or orchestration strategy a general correctness label.

\subsection{System Development Implications}
\label{subsec:discussion-system-implications}

The results have direct implications for the design of LLM-based refactoring systems. Traditional refactoring tools aim to automate behaviour-preserving transformations or help developers apply refactorings with reduced manual effort~\cite{mens2004survey,murphyhill2012how}. Developers nevertheless report that refactoring interacts with integration, review, regression risk, and project-specific constraints~\cite{kim2012field}. These concerns remain important when the transformation is generated by an LLM.

First, systems should perform pre-generation exposure screening. The target file's dependency surface, caller-used APIs, downstream callers, and public or protected interfaces can indicate how much repository-level verification may be needed. The file-level RQ5 analysis shows that exposure measures are associated with High-risk share and broken-caller share. A system could use these measures to select an appropriate verification policy, but not to predict semantic failure. High-exposure files may require stricter API-preservation constraints, smaller transformations, mandatory caller analysis, or earlier human review.

Second, systems should separate local improvement from repository-level verification. Smell reduction, complexity reduction, and maintainability proxies describe the intended benefit of a refactoring. Contract changes, dependency changes, caller impact, behavioural indicators, compilation, and tests describe different aspects of adoption evidence. Combining these measures into one score can conceal important trade-offs. A system should report the intended improvement and the available preservation evidence separately.

Third, systems should report risk mechanisms rather than only an aggregate label. The High-risk category is produced by overlapping structural rules, and its prevalence changes when the dependency threshold changes. A developer needs to know whether a candidate changes a caller-used API, modifies a public signature, removes an API, changes a constructor, exceeds the dependency-delta threshold, affects annotations, or changes inheritance relations. The measured value and the rule that triggers the warning should both be visible.

Fourth, systems should treat generated edits as candidates in a staged verification process. Structural analysis is available for all eligible candidates in this study, but compilation evidence is unavailable and targeted-test evidence is uneven. Only 395 candidates are test eligible, and only 65 produce conclusive targeted-test outcomes. When verification evidence is missing or inconclusive, a system should expose this uncertainty rather than converting unavailable evidence into a Low-risk outcome.

Fifth, systems should distinguish screening from validation. A conservative screening rule can be useful even when it produces many warnings, provided that users understand what those warnings represent. High contract/dependency risk should therefore be presented as a request for further verification, not as a verdict that a candidate is incorrect. Candidates with detected broken callers or test regressions may deserve higher review priority because these indicators provide more direct impact evidence.

Overall, LLM-based refactoring systems should be risk-aware by design. They should screen target files, constrain transformations according to repository exposure, report improvement and preservation evidence separately, expose the mechanisms behind warnings, and require appropriate verification before patch application. This design does not treat LLM-generated refactorings as unusable. It treats them as candidates whose readiness depends on the strength and coverage of the available evidence.

\subsection{Research Implications}
\label{subsec:discussion-research-implications}

The results also have implications for empirical research on LLM-based refactoring and repository-level code generation. Recent work increasingly argues that evaluation should move beyond isolated programming tasks toward repository-level settings, executable tasks, and multiple forms of evidence~\cite{bairi2024codeplan,hai2025impacts,abdollahi2026benchmarking}. This direction is particularly important for refactoring because a candidate is a proposed change to an existing software system rather than an isolated code fragment.

First, future benchmarks should characterize repository exposure as part of the task definition. Files that appear similar in size or smell profile may have different callers, public interfaces, inheritance relationships, and dependency surfaces. Reporting these characteristics would allow researchers to distinguish generation performance from differences in task exposure. It would also support stratified analysis rather than treating all selected files as equally difficult or equally risky.

Second, studies should separate rule-based screening outcomes from observed-impact evidence. A configured High-risk classification, detected broken callers, behavioural-equivalence indicators, compilation outcomes, and test outcomes measure different constructs. They should not be collapsed into a single binary correctness variable. Studies should report how each outcome is produced, its coverage, and whether it contributes to the classification rule or provides separate evidence for assessing the candidate.

Third, future work should estimate the precision and recall of screening indicators. The present study can report how frequently the configured rules trigger, but it cannot determine how many High-risk candidates correspond to confirmed semantic failures. Stratified manual review, independent raters, inter-rater agreement measurement, parser-based validation of extraction rules, and more complete build and test evidence would allow future studies to estimate false-positive and false-negative rates. Such validation is necessary before a rule-based classification can be interpreted as a measure of actual refactoring correctness.

Fourth, threshold sensitivity should become a standard part of rule-based evaluation. In this study, changing the dependency-delta threshold materially changes the number of candidates classified as High risk. Reporting only one threshold can make a policy choice appear to be an empirical property of the generated candidates. Future studies should report results across defensible thresholds, justify the selected operating point, and distinguish continuous measurements from the categories derived from them.

Fifth, future studies should strengthen compilation and behavioural validation. Testing research on refactoring has long emphasized regression testing and impact-aware validation for detecting behaviour changes~\cite{soares2017test,wang2018towards}. For LLM-generated refactorings, this requirement is especially important because a generated edit may combine an intended transformation with unintended structural changes. Future evaluations should aim for uniform compilation, targeted testing, regression testing, and repository integration where infrastructure permits. They should also report infrastructure failures separately from candidate failures.

Sixth, studies comparing prompts, agents, or models should preserve the pairing and clustering structure of the experiment. The present study contains 705 complete eligible pairs from 750 planned file-model pairing keys. Model comparisons also require complete five-model blocks whose sizes differ between generation conditions. Analyses that ignore these structures can overstate precision or compare different populations. File-level aggregation, paired tests, clustered models, or other dependency-aware statistical methods should be selected according to the experimental design.

Finally, future work should investigate exposure-aware generation rather than relying only on post-generation filtering. Low-exposure files may permit a wider range of transformations, while high-exposure files may require stricter API preservation, smaller edits, caller-aware planning, or mandatory human review. Such systems would adapt the generation and verification strategy to the repository role of the target file.

Taken together, these implications support a more evidence-aware research agenda for LLM-based refactoring. Future work should evaluate whether a model improves code locally, how the candidate changes repository-facing structures, what observed impact is detected, and how much independent verification is available. This approach would better reflect how refactorings are reviewed, integrated, and trusted in real software projects.

%% file: sections/06-threats-to-validity.tex
\section{Threats to Validity}
\label{sec:threats}

This section discusses the threats that may affect the validity of the study and the measures taken to reduce their impact. Following established empirical software-engineering practice, we organize the threats into construct, internal, external, and conclusion validity, followed by reliability and reproducibility.

\subsection{Construct Validity}
\label{subsec:construct-validity}

Construct validity concerns whether the measures used in the study adequately represent repository-level risk, behavioural preservation, and adoption readiness.

\textbf{Operationalization of repository-level risk.}
The main construct threat concerns the interpretation of the pipeline-defined High contract/dependency-risk classification. This classification is produced by configured structural rules concerning caller-used APIs, public and protected signatures, constructors, dependencies, annotations, inheritance relations, and dependency-delta magnitude. It is therefore a screening outcome rather than direct evidence of semantic failure. \textit{Mitigation.}
We report High contract/dependency risk separately from detected broken callers, the behavioural-equivalence indicator, compilation evidence, and targeted-test outcomes. We avoid interpreting High risk as confirmed failure or as the probability that a candidate is semantically incorrect. Adoption-readiness categories are presented as screening and review states under the available evidence.

\textbf{Unknown precision of the High-risk classification.}
The study does not contain a sufficiently large independently reviewed sample from which to estimate the precision or false-positive rate of the High-risk classification. Consequently, the observed High-risk prevalence may reflect both genuine structural concerns and the conservativeness of the configured rules. \textit{Mitigation.}
We frame the 93.8\% High-risk result as the frequency with which the implemented screening rules trigger. We do not present it as the proportion of candidates that would fail during integration or execution. We also report more direct evidence, including detected broken callers and targeted-test regressions, as separate outcomes.

\textbf{Dependency-threshold selection.}
The classification of a large dependency delta depends on a threshold of five. This threshold is a policy choice and influences the number of candidates classified as High risk. \textit{Mitigation.}
We conduct a threshold-sensitivity analysis using thresholds of 5, 7, 10, 15, and 20. Within the 864 candidates having usable risk-reason information, the number of High-risk candidates decreases from 810 at a threshold of five to 714 at a threshold of 20. We therefore report the threshold explicitly and avoid treating it as a universal boundary between semantically correct and incorrect refactorings.

\textbf{Approximation of caller impact and behavioural preservation.}
Broken-caller evidence is derived from static analysis of caller references affected by generated changes. The behavioural-equivalence indicator is based on changes to constructs such as exception handling, conditional logic, critical method calls, and framework-related elements. These measures may identify potential impact without demonstrating observable runtime failure. They may also miss behavioural changes that do not alter the inspected structures. \textit{Mitigation.}
We describe broken-caller evidence as detected static impact and the behavioural-equivalence measure as a heuristic indicator. Neither is presented as proof of runtime failure or behavioural non-equivalence. We interpret these measures together with structural, test, and artifact-availability evidence rather than treating any one indicator as complete validation.

\textbf{Partial targeted-test evidence.}
Targeted tests do not provide uniform behavioural evidence across the eligible cohort. Of the 395 test-eligible candidates, only 65 produce conclusive outcomes. Sixty pass the available targeted tests and five exhibit regressions. All test-eligible candidates belong to the multi-agent condition. \textit{Mitigation.}
We state the denominator for each test result and restrict interpretation to the conclusive test subset. We do not generalize the 60 passing outcomes to the complete eligible cohort or compare test outcomes across generation conditions. We also report that 50 of the 60 test-passing candidates receive the High contract/dependency-risk classification. This separation prevents targeted-test success from being interpreted as validation or invalidation of the structural-screening rules.

\textbf{Measurement of local improvement.}
The study uses configured PMD rule violations to represent local refactoring need and smell-related improvement. PMD does not capture every possible design smell or all dimensions of maintainability. \textit{Mitigation.}
We apply the same PMD configuration consistently to source files and generated candidates. We interpret smell reduction as a reduction in configured PMD-detected violations rather than as complete improvement in design quality, readability, or developer-perceived maintainability. Local improvement is also reported separately from repository-level verification evidence.

\subsection{Internal Validity}
\label{subsec:internal-validity}

Internal validity concerns whether the observed outcomes can be attributed to the evaluated generation conditions and analysis procedures rather than uncontrolled factors.

\textbf{Differences between generation conditions.}
The single-prompt and multi-agent conditions differ in more than the number of prompts. The single-prompt condition performs analysis, planning, transformation, and preservation reasoning in one interaction, whereas the multi-agent condition separates these activities across specialized stages. Observed differences may therefore result from orchestration, intermediate instructions, verifier feedback, or output budgets. \textit{Mitigation.}
Both conditions use the same selected files, model configurations, and repository-level dependency context. Generation-condition comparisons use matched candidates for the same project, file, and model. We interpret the results as comparisons between complete generation configurations rather than as causal estimates of the contribution of an individual agent or workflow component.

\textbf{Incomplete generated candidates.}
Some model responses were truncated or did not contain the required primary Java type. Applying candidate-level risk analysis to such outputs could produce misleading classifications because incomplete code may still contain fragments that match structural extraction rules. \textit{Mitigation.}
We apply explicit candidate-eligibility criteria before candidate-level analysis. Of the 1,500 generation attempts, 45 do not produce a complete analyzable Java candidate and are excluded from risk, adoption-readiness, and association analyses. These outcomes remain included in generation-completeness reporting. The analytical cohort therefore contains 1,455 eligible candidates.

\textbf{Unequal distribution of incomplete generations.}
Incomplete generations are not evenly distributed across the experimental conditions. Forty-four incomplete generation-condition pairs result from an ineligible single-prompt candidate, while one results from an ineligible multi-agent candidate. Most incomplete generations also occur in the Qwen configuration. Complete-case comparisons may therefore be affected if generation completeness is associated with model behaviour or task difficulty. \textit{Mitigation.}
Paired generation-condition analyses use only the 705 pairs for which both candidates satisfy the eligibility criteria. Model-level tests use complete model blocks within each generation condition. We report the corresponding denominators and interpret the findings within the analyzed populations rather than extending them to all generation attempts.

\textbf{Lightweight Java-source analysis.}
Repository profiling and candidate comparison rely on regular-expression-based analysis rather than complete semantic Java resolution. This may affect dependencies involving reflection, dynamic dispatch, dependency injection, wildcard imports, overloaded symbols, name collisions, generated sources, or unusual project structures. Caller-used APIs and downstream callers are therefore approximations rather than a precise interprocedural call graph. \textit{Mitigation.}
We apply the same profiling and extraction implementation consistently to all repositories, models, and generation conditions. We report the resulting measures as static structural indicators and avoid presenting them as exact runtime call relationships. Claims based on caller and dependency analysis are limited to the constructs measured by the implemented pipeline.

\textbf{Incomplete risk-reason coverage.}
Machine-readable risk reasons are available for 864 of the 1,455 eligible candidates, including all 749 multi-agent candidates but only 115 of the 706 single-prompt candidates. The covered single-prompt candidates may not represent the complete single-prompt cohort. \textit{Mitigation.}
We restrict mechanism-specific rates to candidates with usable risk-reason information and report the corresponding denominator. Generation-condition comparisons of individual reason tags use only the 115 complete pairs with usable reasons in both conditions. We apply Holm correction across the reason-tag comparisons, after which no reason-tag difference remains statistically significant.

\textbf{Heuristic dependency-risk score.}
The dependency-risk score combines repository-exposure measures using fixed weights. These weights represent the study's screening priorities but have not been validated as a causal model of refactoring failure. \textit{Mitigation.}
We use the score as an operational exposure measure rather than as a predictive failure model. RQ5 also reports its constituent characteristics, including fan-in, fan-out, caller-used APIs, downstream callers, and public or protected API count. This allows the associations to be interpreted without relying solely on the composite score.

\textbf{Model invocation and nondeterminism.}
LLM outputs may vary because of provider routing, prompt wording, context length, output limits, timeouts, stochastic behaviour, and changes to hosted models. The multi-agent workflow also uses different roles and output budgets across its stages. \textit{Mitigation.}
We record the configured model identifiers, prompts, selected files, generation settings, and dependency context used in the experiment. The same task inputs and primary configuration are used across comparable runs. Nevertheless, the study evaluates the observed outputs of the specified configurations and does not claim that rerunning the models would produce identical candidates.

\subsection{External Validity}
\label{subsec:external-validity}

External validity concerns the extent to which the results can be generalized beyond the evaluated repositories, files, models, and generation settings.

\textbf{Selection of subject systems.}
The study uses 15 open-source Java repositories. Although these repositories represent several application domains and structural profiles, they do not represent all Java systems or industrial software. \textit{Mitigation.}
We select multiple repositories rather than relying on a single system and apply the same automated profiling and generation procedures across all projects. The findings are therefore supported across the included systems, but we limit generalization to production Java files with characteristics similar to the selected cohort.

\textbf{Non-random file selection.}
The 150 selected files are not a random sample of all files in the subject repositories. The selection procedure intentionally combines local refactoring need with repository-facing exposure. Files with no fan-in, no fan-out, no public API, and no caller evidence are excluded. \textit{Mitigation.}
We explicitly describe the file-selection criteria and characterize the selected files in RQ1. The reported risk rates are interpreted as results for exposure-bearing production files rather than as prevalence estimates for all Java files or all possible refactoring tasks.

\textbf{Programming-language scope.}
The study focuses exclusively on Java. Java provides explicit packages, imports, classes, access modifiers, inheritance relations, and type dependencies. Repository-level exposure may appear differently in dynamically typed languages, scripting languages, generated-code-heavy systems, or polyglot repositories. \textit{Mitigation.}
We restrict our claims to Java-based refactoring and avoid generalizing the specific indicator frequencies to other programming languages. The repository-level verification concepts may transfer to other ecosystems, but their extraction rules and thresholds require separate evaluation.

\textbf{Model and workflow scope.}
The results are specific to the five evaluated model configurations, provider-routed identifiers, prompts, generation conditions, orchestration design, and verification pipeline. Different models, prompts, retrieval strategies, repair loops, or verification tools may produce different outcomes. \textit{Mitigation.}
We compare multiple model configurations under both generation conditions and report model-specific as well as aggregate results. We frame the conclusions around the evaluated configurations rather than claiming that they apply to every LLM or multi-agent refactoring system.

\textbf{File-level generation scope.}
The study generates replacement candidates for individual files rather than complete pull requests or coordinated multi-file transformations. Some detected caller or contract risks could potentially be resolved through coordinated changes to dependent files. Multi-file generation could also introduce additional risks not captured in this study. \textit{Mitigation.}
We define the candidate and unit of analysis explicitly and limit the conclusions to file-level refactoring generation. Repository-level evidence is used to assess the impact of these file-level candidates, but the results are not generalized to complete automated repository transformation.

\textbf{Open-source setting.}
Industrial repositories may contain proprietary frameworks, generated sources, organization-specific APIs, different architectural constraints, stronger test suites, and different review practices. \textit{Mitigation.}
The selected repositories vary in size, domain, and dependency structure, which provides some diversity within the open-source setting. However, we do not claim direct generalizability to industrial systems. The findings motivate, rather than replace, further evaluation in company-specific environments.

\subsection{Conclusion Validity}
\label{subsec:conclusion-validity}

Conclusion validity concerns whether the statistical procedures and analytical decisions support the conclusions drawn from the data.

\textbf{Dependence among generated candidates.}
Multiple candidates are generated from the same source file, so candidate observations are not fully independent. Treating all candidates as independent could overstate statistical precision. \textit{Mitigation.}
Generation-condition analyses use matched project-file-model pairs. Model comparisons use complete five-model blocks within each generation condition. RQ5 uses the source file as the primary unit for pre-generation association analysis, producing 150 file-level observations. Candidate-level RQ5 analyses are retained only as sensitivity checks. File-level aggregation reduces, but does not eliminate, dependence because files remain clustered within projects.

\textbf{Statistical characteristics of the outcomes.}
Several outcomes are binary, zero-inflated, skewed, or highly imbalanced. Broken-caller counts are zero for most candidates, while a small number of candidates have comparatively large counts. The High-risk classification occurs for 1,365 of the 1,455 eligible candidates, leaving relatively few non-High cases. \textit{Mitigation.}
We use exact McNemar tests for paired binary outcomes, a tie-corrected Wilcoxon signed-rank procedure for paired broken-caller counts, Cochran's $Q$ tests for matched binary model comparisons, Friedman tests for matched count comparisons, Spearman correlations for file-level continuous associations, and Mann--Whitney procedures with Cliff's $\delta$ for binary file characteristics. We report effect sizes, confidence intervals, and outcome-specific denominators alongside statistical significance.

\textbf{Limited statistical power for paired comparisons.}
The paired generation-condition analysis does not identify statistically supported differences in High risk, broken-caller occurrence, a false behavioural-equivalence indicator, or broken-caller count. A non-significant result does not demonstrate that the two conditions are equivalent. \textit{Mitigation.}
We report the observed counts, effect directions, discordant pairs, and exact or adjusted $p$-values rather than interpreting non-significance as equality. We conclude only that the study does not provide evidence of a general generation-condition difference for these outcomes.

\textbf{Multiple comparisons.}
Comparing several models, indicators, and source-file characteristics increases the probability of chance findings. \textit{Mitigation.}
We apply Holm correction within the declared statistical families. Unadjusted values are not used as the sole basis for substantive claims. Statistical results are interpreted together with effect sizes, denominators, data coverage, and consistency across related analyses.

\textbf{Exploratory association analysis.}
RQ5 examines associations between source-file characteristics and three outcomes. These associations could be misinterpreted as causal or predictive relationships. Risk-reason flags also overlap mechanically with the High-risk classification. \textit{Mitigation.}
We use the file level as the primary analytical unit, describe the analysis as exploratory, and avoid causal language. Risk-reason flags are presented as a decomposition of the implemented classification rather than as independent predictors. High contract/dependency risk, broken-caller evidence, and a false behavioural-equivalence indicator are analyzed separately.

\textbf{Missing post-generation telemetry.}
Churn, lines added, and lines removed are available for 778 candidates but missing for 677 of the 749 multi-agent candidates. Available-case analyses for these measures are therefore dominated by single-prompt candidates. \textit{Mitigation.}
Missing telemetry is retained as missing rather than converted to zero. We report the available-case denominator and do not generalize the corresponding findings to the complete eligible cohort. Smell delta, which is available for all eligible candidates, is analyzed separately.

\textbf{Incomplete behavioural and build evidence.}
Compilation evidence is unavailable, and targeted-test evidence is both partial and restricted to the multi-agent condition. These limitations prevent estimation of the diagnostic accuracy of the High-risk classification. \textit{Mitigation.}
We report structural screening, caller impact, the behavioural-equivalence indicator, and targeted-test evidence as separate constructs. Test findings are restricted to the test-eligible and conclusive subsets. The study reports screening prevalence and exploratory associations rather than precision, recall, or real-world failure probability.

\subsection{Reliability and Reproducibility}
\label{subsec:reliability-reproducibility}

Reliability and reproducibility concern whether the study procedures and analyses can be inspected or repeated and whether rerunning the experiment would produce consistent results.

\textbf{Changes to provider-hosted models.}
Provider-routed models and serving infrastructure may change over time, even when the configured model identifier remains the same. Model nondeterminism may also produce different candidates in repeated executions. \textit{Mitigation.}
We record the configured model identifiers, prompts, generation settings, selected files, and dependency context used for each condition. These records support replication of the study configuration, although exact reproduction of generated outputs cannot be guaranteed.

\textbf{Reproducibility of file selection and cohort construction.}
Differences in selected files or candidate-eligibility decisions could affect the analytical cohort. \textit{Mitigation.}
The file-selection procedure is automated and uses a fixed seed. Both generation conditions use the same project-file manifest. Candidate eligibility is determined using explicit completeness criteria, and generation attempts are distinguished from analyzable candidates. This results in 1,500 generation attempts, 1,455 eligible candidates, and 705 complete generation-condition pairs.

\textbf{Incomplete artifact retention.}
Generated candidate bodies and associated artifacts are available for independent inspection for 187 eligible candidates. For the remaining 1,268 candidates, the study outcome records are available, but the generated files cannot be independently reconstructed or reprofiled. \textit{Mitigation.}
We distinguish candidate-level outcome records from candidate artifacts available for direct inspection and report artifact availability as part of the dataset. Conclusions requiring direct candidate inspection are restricted accordingly. The unavailable artifacts remain a limitation on complete independent reproduction.

\textbf{Reproducibility of derived measures.}
Repository profiles, risk reasons, High-risk classifications, adoption-readiness categories, and statistical results depend on the implemented analysis rules and their configuration. \textit{Mitigation.}
We preserve the repository-profile data, candidate-level outcome records, threshold-sensitivity results, analytical denominators, paired-comparison data, file-level aggregates, and statistical-analysis inputs. The methodology describes the dependency-delta rule, risk combination, adoption-readiness mapping, eligibility criteria, and statistical procedures used to derive the reported results.

\textbf{Denominator consistency.}
The study uses several analytical populations, including generation attempts, eligible candidates, complete pairs, reason-covered candidates, test-eligible candidates, conclusive test outcomes, and candidates with artifacts available for independent inspection. Confusing these populations could lead to inconsistent or irreproducible conclusions. \textit{Mitigation.}
We report the relevant denominator beside each result and restrict comparisons to populations for which the required evidence is available. This separation allows the findings to be interpreted according to the evidence supporting each analysis.

%% file: sections/07-conclusion.tex
\section{Conclusion}
\label{sec:conclusion}

This paper investigated repository-level screening and verification evidence for LLM-generated Java refactoring candidates. Across 15 open-source repositories, 150 production files, five LLM configurations, and two dependency-aware generation conditions, we conducted 1,500 generation attempts, of which 1,455 produced complete candidates eligible for analysis.

The results show that local code improvement and repository-level adoption evidence capture different properties. Among the eligible candidates, 1,365 (93.8\%) triggered the pipeline-defined High contract/dependency-risk classification, while 106 (7.3\%) had detected broken-caller evidence. High contract/dependency risk represents a conservative structural-screening outcome rather than confirmed semantic failure, and its prevalence depends partly on the configured dependency-delta threshold. Among the 65 candidates with conclusive targeted-test outcomes, 60 passed, including 50 that also received the High-risk classification. Moreover, the paired analysis of 705 complete generation-condition pairs identified no statistically significant difference between single-prompt and multi-agent generation in the principal risk indicators.

Overall, LLM-generated refactorings should be treated as candidates requiring staged repository-level verification rather than as directly adoptable patches. Future research should validate the precision of structural-screening indicators through stratified manual review, strengthen parser-based and compilation-based verification, expand behavioural testing, examine coordinated multi-file refactoring, and develop exposure-aware generation strategies that adapt refactoring and verification policies to the repository role of the target file.